\documentclass[12pt]{article}
\usepackage[english]{babel}
\usepackage{amsmath}
\usepackage{amssymb}%
\usepackage{cite}

\begin{document}

\title{\hfill{\normalsize{}hep-th/0703049}\\[5mm]
{\bf{} BRST approach to Lagrangian construction for fermionic higher
spin fields in AdS space}}

\author{\sc I.L. Buchbinder${}^{a}$\thanks{joseph@tspu.edu.ru
\hspace{0.5cm} ${}^{\dagger}$krykhtin@mph.phtd.tpu.edu.ru
\hspace{0.5cm} ${}^{\ddagger}$ reshet@tspu.edu.ru}, V.A.
Krykhtin${}^{a,b\dagger}$, A.A.
Reshetnyak$^{c\ddagger}$ \\[0.5cm]
\it ${}^a$Department of Theoretical Physics,\\
\it Tomsk State Pedagogical University,\\
\it Tomsk 634041, Russia\\[0.3cm]
\it ${}^{b}$Laboratory of Mathematical Physics,\\
\it Tomsk Polytechnic University,\\
\it Tomsk 634034, Russia\\[0.3cm]
\it ${}^{c}$Laboratory of Non-equilibrium State Theory,\\
\it  Institute of  Strength Physics
and Materials Science,\\
\it Tomsk 634021, Russia}
\date{}

\maketitle

\begin{abstract}
We develop a general gauge-invariant Lagrangian construction for
half-integer higher spin fields in the AdS space of any dimension.
Starting with a formulation in terms of an auxiliary Fock space, we
obtain closed nonlinear symmetry algebras of higher spin
fermionic fields in the AdS space and find the corresponding BRST
operator. A universal procedure for constructing gauge-invariant
Lagrangians describing the dynamics of fermionic fields
of any spin is developed. No off-shell constraints for the fields
and gauge parameters are imposed from the very beginning. It is
shown that all the constraints determining an irreducible
representation of the AdS group arise  as a consequence of the
equations of motion and gauge transformations. As an example of
the general procedure, we derive gauge-invariant Lagrangians
for massive fermionic fields of spin 1/2 and 3/2 containing the
complete set of auxiliary fields and gauge symmetries.
\end{abstract}

\section{Introduction}\label{Introduction}

Problems of higher spin field theory have attracted much attention
for a long time due to the hope of finding new possibilities and
approaches to the unification of the fundamental interactions.
Higher spin field theory is closely related to superstring theory,
which operates with an infinite tower of bosonic and fermionic
higher spin fields, including massless and massive ones. A Lagrangian
formulation for an interacting higher spin field theory is one
(perhaps the last) of the unsolved general problems of classical
field theory (see, e.g., the reviews \cite{reviews}).

The current studies in higher spin field theory concern the
aspects of Lagrangian formulation in various dimensions, searches
for supersymmetric generalizations, finding a correspondence with
superstring and $M$ theories and a construction of interacting
Lagrangians (see \cite{We}--\cite{Medeiros} for massive and
\cite{Vasiliev}--\cite{Barnich} for massless higher spin
theories).

It is well-known that a space of constant curvature, in particular,
the AdS space, is the simplest non-trivial background providing a consistent
propagation  of higher spin fields (see, e.g., \cite{We,FF}).
We point out two attractive features of higher spin theory in
the AdS space. First, the radius of the AdS space ensures the
presence of a natural dimensional parameter for the accommodation
of compatible self-interactions for massless higher spin fields
\cite{Fradkin,Sezgin}. Second, higher spin fields in the AdS space
are closely related to the conjecture \cite{Maldacena} on the AdS/CFT
correspondence between the conformal $\mathcal{N} = 4$ SYM theory and
the superstring theory on the $AdS_5 \times S_5$ Ramond--Ramond background
\cite{Heslop}.

The study of a Lagrangian formulation for massive higher spin
field theories was initiated by the pioneering works of
Fierz--Pauli and Singh--Hagen \cite{FP} for theories in the
four-dimensional Minkowski space. These works demonstrated a
specific feature of higher spin field theories: all such theories
include, apart from the basic fields with a given spin, also some
auxiliary fields with lesser spins, which provide a compatibility
of the Lagrangian equations of motion with the constraints
determining an irreducible representation of the Poincare group.
Attempts to construct Lagrangian descriptions of free higher spin
field theories in curved spaces and derive higher spin
interactions have resulted in consistency problems which generally
remain open in spite of numerous efforts.

The present work is devoted to a derivation of general
gauge-invariant Lagrangian for massive and massless fields of any
half-integer spin in AdS spaces of arbitrary dimensions. Our
approach is based on use of the BRST--BFV construction \cite{BFV}
(see also the reviews \cite{bf}), which was initially developed
for the quantization of gauge theories. Following a tradition
accepted in string theory and in higher spin field theory, we
further refer to this construction as the BRST construction, and
to the corresponding BFV charge, as the BRST charge. The general
application of the BRST construction to higher spin field theory
consists in three steps. First, the  conditions determining the
representation with a given spin are regarded as first-class
constraints. Second, using the algebra of these constraints, one
constructs the BRST charge. Third, a higher spin field Lagrangian
is constructed in terms of the BRST charge in such a way that the
corresponding equations of motion reproduce the initial
constraints. The final step is of fundamental importance and
ensures the correctness of the Lagrangian construction. We
emphasize that this approach automatically implies a
gauge-invariant Lagrangian. One can expect that in the case of
massive theories a Lagrangian resulting from the BRST construction
should contain a number of St$\ddot{\rm u}$ckelberg fields.

A derivation of Lagrangians for massless bosonic higher spin fields
on a basis of the BRST approach has been considered in \cite{Ouvry,Pashnev}
for the flat space and in \cite{B-BRST-Ads} for the AdS
space. A generalization of the BRST approach to Lagrangians of massless
fermionic higher spin fields in flat spaces of arbitrary dimensions
has been given in \cite{0410215}. An interaction vertex for massless
bosonic higher spin fields in flat and AdS spaces has been studied
within the BRST approach in \cite{0609082}. The first implementation
of the BRST construction for the derivation of Lagrangians for massive
higher spin fields has been given in our papers for bosonic \cite{0505092}
and fermionic \cite{0603212} models in Minkowski spaces of arbitrary
dimensions and for bosonic models in AdS spaces of arbitrary dimensions
\cite{0608005}. Thus, the BRST approach presents a universal
generating method of constructing Lagrangian formulations for
higher spin fields.

In this paper, we develop a gauge-invariant  approach to a
Lagrangian construction for totally symmetric fermionic higher
spin fields in the AdS space of any dimension\footnote{For the
sake of completeness, we note that a Lagrangian formulation for
massive fields of an arbitrary half-integer spin in the AdS space
subject to algebraic constraints has been recently suggested  in
\cite{0609029}.}. The conditions determining an irreducible
representation of the AdS group with a given spin on tensor-spinor
fields are reformulated as operator constraints in an auxiliary
Fock space. These constraints form a nonlinear superalgebra with a
central bosonic charge. Such an algebra has a more complex
structure in comparison with its counterpart for bosonic higher
spin fields in the AdS space \cite{0608005}, due to the presence
of non-vanishing operator structure functions resolving the Jacobi
identity, and therefore the construction of the corresponding BRST
charge faces a problem. Some aspects of the BRST construction for
bosonic nonlinear constraint algebras have been discussed in
\cite{Sevrin,0206027}. We find a solution of the BRST construction
for a nonlinear superalgebra in the theory under consideration and
obtain a Lagrangian which reproduces the initial constraints as a
consequence of the equations of motion.

The paper is organized as follows. In Section~\ref{Fock}, we
examine a closed operator superalgebra based on the constraints
that determine an irreducible representation of the AdS group with
a half-integer spin. In Section~\ref{addparts}, we find a
superalgebra of the additional parts for the  superalgebra of a
modified set of the initial operators, obtained by a linear
transformation of the initial constraints, and then realize its
representation in terms of new (extra) creation and annihilation
operators. In Section~\ref{BRST}, we calculate a deformed
nonlinear superalgebra of constraints enlarged by the additional
parts of the modified constraints. A similar construction for
fermionic fields in flat space has been given in \cite{0410215}.
Next, we construct a BRST operator corresponding to the
superalgebra of the modified enlarged constraints. The derivation
of an action and of a sequence of reducible gauge transformations
describing the propagation of a fermionic field of an arbitrary
spin in the AdS space is realized in Section~\ref{Lagr-constr}. In
Section~\ref{reduction}, we prove that the constructed action
reproduces the correct conditions for the field that determine an
irreducible representation of the AdS group with a fixed
half-integer spin. In Section~\ref{examples}, we illustrate the
procedure by constructing gauge-invariant Lagrangians for the
fields of spin $1/2$ and $3/2$. In Conclusion, we summarize the
results of the work and discuss some open problems.


In addition to the conventions of Refs. \cite{0603212, 0608005},
we use the notation $\varepsilon(A)$, $gh(A)$ for the respective
values of Grassmann parity and ghost number of a quantity $A$, and
denote by $[A,\,B\}$ the supercommutator of quantities $A, B$,
which in the case of definite values of Grassmann parity is given by
$[A\,,B\}$ = $AB - (-1)^{\varepsilon(A)\varepsilon(B)}BA$.

\section{Auxiliary Fock space for higher spin fields in AdS
space-time}\label{Fock}

The massive half-integer spin $s=n+\frac{1}{2}$ representations of
the AdS group are realized in the space of totally symmetric
tensor-spinor fields $\Phi_{\mu_1\ldots\mu_n}(x)$, the Dirac index
being suppressed, satisfying to the conditions (see e.g.
\cite{Metsaev-m})
\begin{eqnarray}
\label{Eq-0} &&
\bigl[i\gamma^{\mu}\nabla_{\mu}
    -r^\frac{1}{2}(n+\textstyle\frac{d}{2}-2)-m
\bigr] \Phi_{\mu_1\mu_2\ldots\mu_n}(x) =0,
\\
 &&
\gamma^{\mu} \Phi_{\mu\mu_2\ldots\mu_n}(x) =0.
 \label{Eq-1}
\end{eqnarray}
where $r=\frac{R}{d(d-1)}$, with $R$ being the scalar curvature of
the space-time. We use the metric with the mostly minus signature,
and the Dirac's matrices obey the relation
\begin{eqnarray}
\bigl\{\gamma_\mu,\gamma_\nu\bigr\}=2g_{\mu\nu}
.
\end{eqnarray}

In order to avoid an explicit manipulation with the indices, it is
convenient to introduce an auxiliary Fock space\footnote{For more details,
see \cite{0608005}.} $\mathcal{H}$ generated by
creation and annihilation operators with tangent space indices
($a,b=0,1,\ldots,d-1$)
\begin{eqnarray}
[a_a,a^+_b]=-\eta_{ab}, \qquad
\eta_{ab}=\mathrm{diag}(+,-,\ldots,-).
\end{eqnarray}
An arbitrary vector in this Fock space has the form
\begin{eqnarray}
\label{PhysState} |\Phi\rangle &=&
\sum_{n=0}^{\infty}\Phi_{a_1\ldots\,a_n}(x)\,
a^{+a_1}\ldots\,a^{+a_n}|0\rangle =
\sum_{n=0}^{\infty}\Phi_{\mu_1\ldots\,\mu_n}(x)\,
a^{+\mu_1}\ldots\,a^{+\mu_n}|0\rangle ,
\end{eqnarray}
where
$a^{+\mu}(x)=e^\mu_a(x)a^{+a}$, $a^\mu(x)=e^\mu_a(x)a^a$;
$e^\mu_a(x)$ being the vielbein.
It is evident that
\begin{eqnarray}
[a_\mu, a_\nu^+]=-g_{\mu\nu}.
\end{eqnarray}
We refer to the vector (\ref{PhysState}) as the basic vector. The
fields $\Phi_{\mu_1\ldots\mu_n}(x)$ are the coefficient functions
of the vector $|\Phi\rangle$ and its symmetry properties are
stipulated by the symmetry properties of the product of the creation
operators. We also assume the standard relation
\begin{eqnarray}
\nabla_\mu e^a_\nu = \partial_\mu e^a_\nu
  -\Gamma_{\mu\nu}^\lambda e^a_\lambda
  +\omega_\mu{}^a{}_b e^b_\nu
=0.
\end{eqnarray}

We intend to realize relations (\ref{Eq-0}), (\ref{Eq-1}) as
certain constraints on the vector $|\Phi\rangle$
(\ref{PhysState}). To this end, we define an operator $D_\mu$
acting on the vectors $|\Phi\rangle$,
\begin{eqnarray}
D_\mu=\partial_\mu-\omega_\mu^{ab}(a_a^+a_b
-\textstyle\frac{1}{4}\gamma_{ab}),
\label{D_mu}
&\qquad&
\partial_\mu|0\rangle=0
,
\qquad
\gamma_{ab}={\textstyle\frac{1}{2}}(\gamma_a\gamma_b-\gamma_b\gamma_a)\,,
\end{eqnarray}
and operators
\begin{eqnarray}
\label{tt0}
\tilde{{t}}_0
&=& i{\gamma}^\mu D_\mu - m - r^{\frac{1}{2}}
(g_0 - 2)
,
\\
\label{tt1}
\tilde{t}_1
&=&
{\gamma}^\mu a_\mu,
\\
g_0&=&-a^{+\mu}a_\mu+\frac{d}{2}
.
\end{eqnarray}
We can see that the constraints
\begin{eqnarray}
\tilde{t}_0|\Phi\rangle=
\tilde{t}_1|\Phi\rangle=0
\label{Eq-2}
\end{eqnarray}
for the basic vector (\ref{PhysState}) are equivalent to equations
(\ref{Eq-0}), (\ref{Eq-1}), with each component
$\Phi_{\mu_1\ldots\mu_n}(x)$ in (\ref{PhysState}) subject to
(\ref{Eq-0}), (\ref{Eq-1}), and therefore they describe a field of spin $n+1/2$.

Because of the fermionic nature of equations (\ref{Eq-0}),
(\ref{Eq-1}) with respect to the standard  Grassmann parity, and
due to the bosonic nature of the operators  $\tilde{t}_0, \tilde{t}_1$:
$\varepsilon(\tilde{t}_0) = \varepsilon(\tilde{t}_1)= 0$,
in order to equivalently transform these operators into
fermionic ones, we now introduce a  set of $d+1$ Grassmann-odd
``gamma-matrix-like objects'' $\tilde{\gamma}^\mu$, $\tilde{\gamma}$
subject to\footnote{For more details, see \cite{0603212}.}
\begin{eqnarray}
\{\tilde{\gamma}^\mu,\tilde{\gamma}^\nu\}
=
2g^{\mu\nu},
\qquad
\{\tilde{\gamma}^\mu,\tilde{\gamma}\}=0,
\qquad
\tilde{\gamma}^2=-1
\label{tgammas}
\end{eqnarray}
and related  to the conventional gamma-matrices as follows:
\begin{eqnarray}\label{gammas}
\gamma^{\mu} = \tilde{\gamma}^{\mu} \tilde{\gamma}.
\end{eqnarray}
In terms of these gamma-matrices, we define the Grassmann-odd
constraints
\begin{eqnarray}
\label{t't0}
{\tilde{t}}'_0
=
-i\tilde{\gamma}^\mu D_\mu +
\tilde{\gamma}\left( m + r^{\frac{1}{2}} (g_0 - 2)\right)
,\qquad
{t}_1 =  \tilde{\gamma}^\mu a_\mu
,
\end{eqnarray}
related to (\ref{tt0}), (\ref{tt1}) as follows:
\begin{eqnarray}
{\tilde{t}}'_0=-\tilde{\gamma}\tilde{t}_0
,
\qquad
t_1=\tilde{\gamma}\tilde{t}_1
.
\end{eqnarray}

In order to obtain a Hermitian BRST operator, being the central
object of the Lagrangian construction in the BRST approach,
we need a set of first-class constraints which must be invariant
with respect to Hermitian conjugation and must form a
superalgebra with respect to supercommutator multiplication. To
this end, we define an odd scalar product:
\begin{equation}
\label{sproduct} \langle\tilde{\Psi}|\Phi\rangle = \int d^dx
        \sqrt{|g|} \sum_{n,\,k=0}^{\infty}
         \langle0|a^{\nu_1}\ldots\,a^{\nu_{k}}
         \Psi^+_{\nu_1\ldots\,\nu_k}(x)
         \tilde{\gamma}_0
         \Phi_{\mu_1\ldots\,\mu_n}(x)\,a^{+\mu_1}\ldots\,a^{+\mu_n}
         |0\rangle
.
\end{equation}
The operators ${\tilde{t}}'_0$, ${t}_1$ (\ref{t't0}) and ${t}_1^+
= \tilde{\gamma}^\mu{}a^+_\mu$, being the Hermitian conjugate of
${t}_1$ with respect to the scalar product (\ref{sproduct}),
generate an operator superalgebra with the central charge $\tilde{m} =
(m -2r^{\frac{1}{2}})$ and the following set of operators:
\begin{align}
&{\tilde{t}}'_0 = -i\tilde{\gamma}^\mu D_\mu +
\tilde{\gamma}\left( m + r^{\frac{1}{2}} (g_0 - 2)\right) \,,
\label{t'0}
\\
& t_1=\tilde{\gamma}^\mu a_\mu\,, \label{t1} &&
t_1^+=\tilde{\gamma}^\mu a_\mu^+
\,,
\\
& l_1=-ia^\mu D_\mu \,, \label{l1} && l_1^+=-i a^{+\mu}D_\mu \,,
\\
& l_2={\textstyle\frac{1}{2}}\,a^\mu a_\mu\,, \label{l2} &&
l_2^+={\textstyle\frac{1}{2}}\,a^{+\mu}a^+_\mu \,,
\\
&&& g_0=-a^+_\mu a^\mu+{\textstyle\frac{d}{2}} \label{g0} ,
\end{align}
\vspace{-2em}
\begin{eqnarray}
{\tilde{l}}'_0=g^{\mu\nu}(D_\nu
D_\mu-\Gamma^\sigma_{\mu\nu}D_\sigma)
-r\left(g_0+t_1^+t_1+{\textstyle\frac{d(d-3)}{4}}\right) + \left(m
+ r^{\frac{1}{2}} (g_0 - 2)\right)^2\,, \label{l'l0}
\end{eqnarray}
which is invariant under Hermitian conjugation. The operators
(\ref{t'0})--(\ref{l'l0}) form an algebra given by
Table~\ref{table in},
\begin{table}[t]
\begin{eqnarray*}
\begin{array}{||c||c|c|c|c|c|c|c|c||c|c||}\hline\hline\vphantom{\biggm|}\hspace{-0.3em}
[\;\downarrow\;,\to\}\hspace{-0.4em}&{\tilde{t}}'_0&t_1&t_1^+&\quad{}{\tilde{l}}'_0&l_1&l_1^+&l_2
&l_2^+ &g_0& \tilde{m}\\
\hline\hline\vphantom{\biggm|} {\tilde{t}}'_0
   &-2{\tilde{l}}'_0&(\ref{t'0t_1})&(\ref{t'0t_1+})&0&(\ref{t'0l1})&-(\ref{l1+t'0})&\hspace{-0.2em}
   -2\tilde{\gamma}r^{\frac{1}{2}}
l_2 \hspace{-0.3em}&\hspace{-0.2em}2\tilde{\gamma}r^{\frac{1}{2}}
 l_2^+ \hspace{-0.2em}&0&0\\
\hline\vphantom{\biggm|} t_1
   &(\ref{t'0t_1})&4l_2&-2g_0&(\ref{t1l'0})&0&-(\ref{l1+t1'})&0&-t_1^+&t_1 & 0\\
\hline\vphantom{\biggm|} t_1^+
   &(\ref{t'0t_1+})&-2g_0&4l_2^+&-(\ref{l'0t1+})&(\ref{l1+t1'})&0&t_1&0&-t_1^+ &0\\
\hline\vphantom{\biggm|} {\tilde{l}}'_0
   &0&-(\ref{t1l'0})&(\ref{l'0t1+})&0&-(\ref{l1l'0})&(\ref{l'0l1+})&-(\ref{l'0l_2})&
   (\ref{l'0+l_2})&0&0\\
\hline\vphantom{\biggm|} l_1
   &-(\ref{t'0l1})&0&-(\ref{l1+t1'})&(\ref{l1l'0})&0&(\ref{l'1l'1+})&0&-l_1^+&l_1 &0 \\
\hline\vphantom{\biggm|} l_1^+
   &(\ref{l1+t'0})&(\ref{l1+t1'})&0&-(\ref{l'0l1+})&-(\ref{l'1l'1+})&0&l_1&0&-l_1^+ & 0\\
\hline\vphantom{\biggm|} l_2
   &\hspace{-0.2em}2\tilde{\gamma}r^{\frac{1}{2}} l_2\hspace{-0.3em}&0&-t_1&(\ref{l'0l_2})&0&-l_1&0&g_0&2l_2&0\\
\hline\vphantom{\biggm|} l_2^+
   &\hspace{-0.2em}-2\tilde{\gamma}r^{\frac{1}{2}}
 l_2^+\hspace{-0.3em}&t_1^+&0&-(\ref{l'0+l_2})&l_1^+&0&-g_0&0&-2l_2^+&0\\
\hline\hline\vphantom{\biggm|} g_0
   &0&-t_1&t_1^+&0&-l_1&l_1^+&-2l_2&2l_2^+&0&0\\
\hline\vphantom{\biggm|} \tilde{m}
   &0&0&0&0&0&0&0&0&0&0\\
   \hline\hline
\end{array}
\end{eqnarray*}
\caption{The algebra of the initial operators.}\label{table in}
\end{table}
where
\begin{eqnarray}
{} [{\tilde{t}}'_0, t^+_1\} & = &   2l^+_1 + \tilde{\gamma}r^{\frac{1}{2}}t^+_1\,,
\label{t'0t_1+}\\
{} [{\tilde{t}}'_0, t_1\} & =  & 2l_1 -
\tilde{\gamma}r^{\frac{1}{2}}t_1\,,\label{t'0t_1}\\
{} [t^+_1\,,l_1\} & = & [l^+_1,t_1\} = {\tilde{t}}'_0 -
\tilde{\gamma}(\tilde{m}+r^{\frac{1}{2}}
 g_0) \,,\label{l1+t1'}
\\
{}[{\tilde{t}}'_0\,,l_1\}&=&r(2t_1^+l_2+g_0t_1-{\textstyle\frac{1}{2}}t_1
)- \tilde{\gamma}r^{\frac{1}{2}}l_1 \label{t'0l1} ,
\\{}
[l_1^+\,,{\tilde{t}}'_0\}&=&r
(2l_2^+t_1+t_1^+g_0-{\textstyle\frac{1}{2}}t_1^+
)-\tilde{\gamma}r^{\frac{1}{2}}l^+_1\label{l1+t'0} ,
\\{}
[t_1\,,{\tilde{l}}'_0\}&=& r(4t_1^+l_2+2g_0t_1-t_1)-
r^{\frac{1}{2}}\tilde{m}t_1\,, \label{t1l'0}
\\{}
[{\tilde{l}}'_0\,,t_1^+\}&=& 4r
(l_2^+t_1+t_1^+g_0)+2r^{\frac{1}{2}}\tilde{m}t_1^+ \label{l'0t1+}
,
\\{}
[l_1\,,{\tilde{l}}'_0\}&=&
4r(l_1^+l_2+g_0l_1)+2r^{\frac{1}{2}}\tilde{m}l_1 \label{l1l'0} ,
\\{}
[{\tilde{l}}'_0\,,l_1^+\}&=&4r(l_2^+l_1+l_1^+g_0) +
2r^{\frac{1}{2}}\tilde{m}l_1^+ \label{l'0l1+} ,
\\{} [l_2\,,{\tilde{l}}'_0\}  & = &
4r^{\frac{1}{2}}(r^{\frac{1}{2}}(1+g_0)+ \tilde{m})l_2\,,
\label{l'0l_2}\\
{}[{\tilde{l}}'_0\,,l^+_2\} & = &
4r^{\frac{1}{2}}l^+_2(r^{\frac{1}{2}}(1+g_0)+ \tilde{m})\,,
\label{l'0+l_2} \\
{} [l_1\,,l_1^+\}&=&{\tilde{l}}'_0 - \tilde{m}^2 -
2r^{\frac{1}{2}}\tilde{m}g_0 +
r({\textstyle\frac{3}{2}}\,t_1^+t_1-{\textstyle\frac{1}{2}}\,g_0-4l_2^+l_2
) \label{l'1l'1+} .
\end{eqnarray}
We call this algebra the massive half-integer higher spin
symmetry algebra in the AdS space.

The method of Lagrangian construction requires an enlarging of the
initial constraints $o_i$ (${\tilde{t}}'_0, {t}_1$ $\in$
$\{o_i\}$), so that the enlarged Hermitian operators contain
arbitrary parameters and the set of enlarged operators form an
(super)algebra. A procedure of constructing of these enlarged
constraints $O_i=o_i+o_i'$ for the operators $o_i$ is considerably
simplified if the initial operators $o_i$ (super)commute with its
additional parts $o_i'$: $[o_i,o_j'\}=0$, defined in the auxiliary
operator space. In this case we can apply the method elaborated in
\cite{0608005}. The such realization of the enlarged first-class
constraints $O_i$, by means of an additive composition of the
quantities $o_i$ with additional parts $o'_i$, requires a
transformation of $o_i$ into an equivalent set of constraints
$\tilde{o}_i$ which do not contain the $\tilde{\gamma}$-matrix,
because the set $o'_i$ will contain this object by  construction.
Thus, in order to simplify the subsequent calculations, we choose
the  operators $\tilde{o}_i$ related by a nondegenerate linear
transformation\footnote{Note that the case of a two-parametric set
of constraints determined by the operator ${\tilde{t}}'_{0\alpha}
= -i\tilde{\gamma}^\mu D_\mu +\alpha_1\tilde{\gamma}\tilde{m}
+\alpha_2\tilde{\gamma}r^{\frac{1}{2}}g_0$, analogous to the
consideration for bosonic fields in the AdS space \cite{0608005},
from which it would be possible to construct a Lagrangian that
reproduces Eq.~(\ref{Eq-2}) for any values $\alpha_1$, $\alpha_2$,
can be realized by means of the above nondegenerate transformation
of the superalgebra of the initial constraints.}
 ${o}_i =U^i_j\tilde{o}_j$
with the initial constraints reproducing relations (\ref{Eq-0}),
(\ref{Eq-1}),
\begin{eqnarray}
t_0 &=&-i\tilde{\gamma}^\mu D_\mu \,, \label{t-0}\\
 l_0 &= &
g^{\mu\nu}(D_\nu D_\mu-\Gamma^\sigma_{\mu\nu}D_\sigma)
-r\left(g_0+t_1^+t_1+{\textstyle\frac{d(d-3)}{4}}\right)
\label{l-0}\,.
\end{eqnarray}
The other constraints coincide with the initial ones.

The operators $\tilde{o}_i$ given by (\ref{t1})--(\ref{g0}),
(\ref{t-0}),(\ref{l-0}) with  the central charge $\tilde{m}$ form
a superalgebra  given by Table~\ref{til-table},
\begin{table}[t]
\begin{eqnarray*}
\begin{array}{||c||c|c|c|c|c|c|c|c||c|c||}\hline\hline\vphantom{\biggm|}
[\;\downarrow\;,\to\}&t_0&t_1&t_1^+&\quad{}l_0&l_1&l_1^+&l_2&l_2^+&g_0&\tilde{m}\\
\hline\hline\vphantom{\biggm|}
t_0
   &-2l_0&2l_1&2l_1^+&0&(\ref{t0l1})&-(\ref{l1+t0})&0&0&0&0\\
\hline\vphantom{\biggm|}
t_1
   &2l_1&4l_2&-2g_0&(\ref{t1l0})&0&-t_0&0&-t_1^+&t_1&0\\
\hline\vphantom{\biggm|}
t_1^+
   &2l_1^+&-2g_0&4l_2^+&-(\ref{l0t1+})&t_0&0&t_1&0&-t_1^+&0\\
\hline\vphantom{\biggm|}
l_0
   &0&-(\ref{t1l0})&(\ref{l0t1+})&0&-(\ref{l1l0})&(\ref{l0l1+})&0&0&0&0\\
\hline\vphantom{\biggm|}
l_1
   &-(\ref{t0l1})&0&-t_0&(\ref{l1l0})&0&(\ref{l1l1+})&0&-l_1^+&l_1 &0 \\
\hline\vphantom{\biggm|}
l_1^+
   &(\ref{l1+t0})&t_0&0&-(\ref{l0l1+})&-(\ref{l1l1+})&0&l_1&0&-l_1^+ & 0\\
\hline\vphantom{\biggm|}
l_2
   &0&0&-t_1&0&0&-l_1&0&g_0&2l_2&0\\
\hline\vphantom{\biggm|}
l_2^+
   &0&t_1^+&0&0&l_1^+&0&-g_0&0&-2l_2^+&0\\
\hline\hline\vphantom{\biggm|}
g_0
   &0&-t_1&t_1^+&0&-l_1&l_1^+&-2l_2&2l_2^+&0&0\\
\hline\vphantom{\biggm|} \tilde{m}
   &0&0&0&0&0&0&0&0&0&0\\
   \hline\hline
\end{array}
\end{eqnarray*}
\caption{The algebra of the modified initial
operators.}\label{til-table}
\end{table}
where
\begin{eqnarray}
[t_0\,,l_1\}&=&r(2t_1^+l_2+g_0t_1-{\textstyle\frac{1}{2}}t_1)
\label{t0l1} ,
\\{}
[t_1\,,l_0\}&=& r(4t_1^+l_2+2g_0t_1-t_1) \label{t1l0} ,
\\{}
[l_1\,,l_0\}&=& r(4l_1^+l_2+2g_0l_1-l_1) \label{l1l0} ,
\\{}
[l_1^+\,,t_0\}&=&r
(2l_2^+t_1+t_1^+g_0-{\textstyle\frac{1}{2}}t_1^+) \label{l1+t0} ,
\\{}
[l_0\,,t_1^+\}&=& r (4l_2^+t_1+2t_1^+g_0-t_1^+) \label{l0t1+} ,
\\{}
[l_0\,,l_1^+\}&=&r(4l_2^+l_1+2l_1^+g_0-l_1^+) \label{l0l1+} ,
\\{}
[l_1\,,l_1^+\}&=&l_0+
r(g_0^2-{\textstyle\frac{1}{2}}\,g_0-4l_2^+l_2
 +{\textstyle\frac{3}{2}}\,t_1^+t_1)
\label{l1l1+}
.
\end{eqnarray}
 In terms of the operators $\tilde{o}_i$, equations (\ref{Eq-2}) have the form
\begin{eqnarray}
&&
\bigl[
t_0+\tilde{\gamma}m+\tilde{\gamma}r^{\frac{1}{2}}(g_0-2)
\bigr]|\Phi\rangle=0,
\qquad
t_1|\Phi\rangle=0.
\label{TheEq}
\end{eqnarray}

In what follows, we shall demonstrate the construction of
Lagrangians by using the BRST approach developed for bosonic
fields in the AdS space in \cite{0608005} so as to reproduce
equations (\ref{Eq-2}), or, equivalently, (\ref{TheEq}). According
to our approach, we have to extend the operators $\tilde{o}_i$ of
the algebra given by Table~\ref{til-table} by additional parts
$o_i'$: $\tilde{o}_i\to\tilde{O}_i=\tilde{o}_i+o_i'$, so that
1)~$\tilde{O}_i$ must be in involution
$[\tilde{O}_i,\tilde{O}_j\}\sim{}\tilde{O}_k$ and 2)~each
Hermitian operator must contain linearly an arbitrary parameter
whose values are to be determined later (for details, see
\cite{0410215, 0603212, 0608005}).

The next step of this procedure is to find the additional parts
$o_i'$ for the operators (\ref{t1})--(\ref{g0}), (\ref{t-0}), (\ref{l-0}).

\section{Additional parts of operators}\label{addparts}

As stated at the end of the previous section, now we intend to
find the additional parts $o_i'$ for the operators
(\ref{t1})--(\ref{g0}), (\ref{t-0}), (\ref{l-0}). Following the
procedure described in the bosonic case in \cite{0608005}, we must
first find the superalgebra of the additional parts. Then these
additional parts are constructed from new (additional) creation
and annihilation operators, as well as from the constants of the
theory $r$, $m$, and from one of the gamma-matrix-like objects
$\tilde{\gamma}$. Furthermore, we must introduce
linearly\footnote{We have to introduce linearly some arbitrary
constants into all of the additional parts corresponding to
Hermitian operators. Since $t_0^2=-l_0$, we cannot obtain an
independent arbitrary constant in $l_0'$.} into $t_0'$, $g_0'$
(these are the additional parts for $t_0$, $g_0$, respectively)
some arbitrary constants whose values will be determined later by
the condition that equations (\ref{TheEq}) be reproduced.

Let us briefly remind and at the same time generalize the method
given in \cite{0608005} to algebras including fermionic operators.
The (super)commutation relations for the operators $\tilde{o}_i$
from Table~\ref{til-table} have the structure
\begin{eqnarray}
[\,\tilde{o}_i,\tilde{o}_j\}_s &=&
f_{ij}^k\tilde{o}_k+f_{ij}^{km}\tilde{o}_k\tilde{o}_m ,
\label{inal}
\end{eqnarray}
where the constants $f_{ij}^k$, $f_{ij}^{km}$ obey the properties
$(f_{ij}^k, f_{ij}^{km})$ =
$-(-1)^{\varepsilon(O_i)\varepsilon(O_j)}(f_{ji}^k, f_{ji}^{km})$.
Then we suppose that the operators $\tilde{o}_i$ supercommute with
the additional parts $o_i'$. In this case, one can check that if
we define the superalgebra of the additional parts in the form
\begin{eqnarray}
[\,o_i',o_j'\}_s &=& f_{ij}^ko_k'
-(-1)^{\varepsilon(\tilde{o}_k)\varepsilon(\tilde{o}_m)}f_{ij}^{km}
o_m'o_k' \,, \label{addal}
\end{eqnarray}
then the enlarged operators $\tilde{O}_i=\tilde{o}_i+o_i'$ will
form a closed superalgebra
\begin{eqnarray}
[\,\tilde{O}_i,\tilde{O}_j\}_s&=& f_{ij}^k\tilde{O}_k
-\left(f_{ij}^{mk}+(-1)^{\varepsilon(O_k)\varepsilon(O_m)}f_{ij}^{km}\right)
o_m'\tilde{O}_k +
f_{ij}^{km}\tilde{O}_k\tilde{O}_m\label{alg-enl},
\end{eqnarray}
which is deformed in comparison with (\ref{inal}).

After applying of the  procedure above, the superalgebra of the
additional parts takes the form given by Table~\ref{table'},
\begin{table}
\begin{eqnarray*}
&
\begin{array}{||c||r|r|r|r|r|r|r|r||r|r||}\hline\hline\vphantom{\biggm|}
       [\;\downarrow\;,\to\}&t'_0&t'_1&t^{\prime+}_1
       &l'_0&l'_1&l^{\prime+}_1&l'_2& l^{\prime+}_2 &g'_0\\
\hline\hline\vphantom{\biggm|}
       t'_0
       &-2{l}'_0& 2l_1'& 2l_1^{\prime +} &0&-(\ref{l1't0'})
       & (\ref{t0'l1'+})& 0& 0 &0\\
\hline\vphantom{\biggm|}
       t'_1
       &2l_1^{\prime } &4l'_2&-2g'_0& -(\ref{l0't1'})&0&
       -{t}'_0 & 0 & -t^{\prime+}_1     &t'_1\\
\hline\vphantom{\biggm|}
       t^{\prime+}_1
          & 2l_1^{\prime +}&-2g'_0&4l^{\prime+}_2&(\ref{t1'+l0'})&
          t'_0 & 0 &t'_1 & 0  &-t^{\prime+}_1 \\
\hline\vphantom{\biggm|}
       l'_0
          &0&(\ref{l0't1'})&-(\ref{t1'+l0'})&0&(\ref{l0'l1'})
          &-(\ref{l1'+l0'}) &  0 &0 &0\\
\hline\vphantom{\biggm|}
       l'_1
       &(\ref{l1't0'})& 0 &-t'_0&-(\ref{l0'l1'})
       &0&(\ref{l1'l1'+}) & 0 &-l^{\prime+}_1 &l'_1 \\
\hline\vphantom{\biggm|}
       l^{\prime+}_1
       &-(\ref{t0'l1'+})&t'_0& 0&(\ref{l1'+l0'})
       &-(\ref{l1'l1'+})& 0 & l'_1 &0&-l^{\prime+}_1\\
\hline\vphantom{\biggm|}
       l'_{2}
       &0&0&-t'_1& 0 & 0 & -l'_1 & 0 & g'_0 &2l'_2\\
\hline\vphantom{\biggm|}
       l^{\prime+}_{2}
       & 0 & t^{\prime+}_1 & 0 &0& l^{\prime+}_1 & 0
       & -g'_0 &0&-2l^{\prime+}_2\\
\hline\hline\vphantom{\biggm|}
       g'_{0}
       &0&-t'_1&t^{\prime+}_1&0&-l'_1&l^{\prime+}_1 &-2l'_2&2l^{\prime+}_2&0\\
       \hline\hline
\end{array}
\end{eqnarray*}
\caption{The algebra of the additional parts for the operators.}
\label{table'}
\end{table}
%
%
%
%
%
where
\begin{eqnarray}
&& [l'_1,t'_0\} =
r(2t^{\prime+}_1l'_2+g'_0t'_1-\textstyle\frac{1}{2}t'_1) ,
\label{l1't0'}
\\
&& [t'_0,l^{\prime +}_1\} = r(2l^{\prime
+}_2t'_1+t^{\prime+}_1g'_0-\textstyle\frac{1}{2}t^{\prime+}_1)
\label{t0'l1'+} ,
\\
&& [l'_0,t'_1\} = r(4t^{\prime+}_1l'_2+2g'_0t'_1-t'_1) ,
\label{l0't1'}
\\
&& [t^{\prime+}_1,l'_0\} =
r(4l^{\prime+}_2t'_1+2t^{\prime+}_1g'_0-t^{\prime+}_1),
\label{t1'+l0'}
\\
&& [l'_0,l'_1\} = r(4l^{\prime +}_1l'_2+2g'_0l_1'-l'_1) ,
\label{l0'l1'}
\\
&& [l^{\prime +}_1,l'_0\} =
r(4l^{\prime+}_2l'_1+2l^{\prime+}_1g'_0-l^{\prime+}_1) ,
\label{l1'+l0'}
\\
&& [l'_1,l^{\prime +}_1\} = l'_0 - r ( g^{\prime{}2}_0
-{\textstyle\frac{1}{2}}g'_0 -4l^{\prime+}_2l'_2
+{\textstyle\frac{3}{2}} t^{\prime +}_1t'_1 ). \label{l1'l1'+}
\end{eqnarray}
In accordance with our method, we ascribe the following value
to the additional central charge: $\tilde{m}' = - \tilde{m}$,
so that the enlarged central charge $\tilde{M}$ becomes equal
to zero.

 Explicit expressions for the additional parts can be
found by the method described in the papers \cite{0410215,
0206027} and extended to the case of the Verma module construction
for a nonlinear superalgebra given by Table~\ref{table'}. Omitting
tedious calculations, we have
\begin{align}
\label{t1'+}
&t_1^{\prime+} = f^+ +2b_2^+f,
&&l_1^{\prime+}=m_1 b_1^+,
\\
\label{g0'}
& g_0'= b_1^+b_1+2b_2^+b_2+f^+f+h,
&& l_2^{\prime+} =b_2^+,
\end{align}
\vspace{-2ex}
\begin{eqnarray}
t_0'
&=&
2m_1b_1^+f
-\frac{m_1}{2}(f^+{-}2b_2^+f)\,b_1^+
    \sum_{k=1}^{\infty}
    \left(\frac{-2r}{m_1^2}\right)^{k}
    \frac{(b_2^+)^{k-1}b_1^{2k}}{(2k)!}
+\tilde{\gamma}m_0\sum_{k=0}^{\infty}
    \left(\frac{-2r}{m_1^2}\right)^{k}
    \frac{(b_2^+)^{k}b_1^{2k}}{(2k)!}
\nonumber
\\
&&{}
+\frac{r(h-{\textstyle\frac{1}{2}})}{m_1}(f^+{-}2b_2^+f )
    \sum_{k=0}^{\infty}\left(\frac{-2r}{m_1^2}\right)^k
    \frac{(b_2^+)^{k}\;b_1^{2k+1}}{(2k+1)!}
,
\label{t0'}
\end{eqnarray}
\vspace{-2ex}
\begin{eqnarray}
t_1'
&=&
-2 g_0'f
-(f^+{-}2b_2^+f)b_2
+\frac{1}{2}(h-{\textstyle\frac{1}{2}})(f^+{-}2b_2^+f)
    \sum_{k=1}^{\infty}
    \left(\frac{-2r}{m_1^2}\right)^{k}
    \frac{(b_2^+)^{k-1}b_1^{2k}}{(2k)!}
\nonumber
\\
&&{}
+\frac{1}{2}(f^+{-}2b_2^+f)\;b_1^+
    \sum_{k=1}^{\infty}
    \left(\frac{-2r}{m_1^2}\right)^{k}
    \frac{(b_2^+)^{k-1}\;b_1^{2k+1}}{(2k+1)!}
\nonumber
\\
&&{}
-\frac{\tilde{\gamma}m_0}{m_1}\;\sum_{k=0}^{\infty}
    \left(\frac{-2r}{m_1^2}\right)^{k}
    \frac{(b_2^+)^{k}\; b_1^{2k+1}}{(2k+1)!}
\label{t1'}
,
\end{eqnarray}
\vspace{-2ex}
\begin{eqnarray}
l_0'
&=&
m_0^2
-r\frac{\tilde{\gamma}m_0}{m_1}\;(f^+{-}2b_2^+f)
    \sum_{k=1}^{\infty}
    \left(\frac{-8r}{m_1^2}\right)^{k}
    \frac{(b_2^+)^k\;b_1^{2k+1}}{(2k+1)!}\;(1-4^{-k})
\nonumber
\\
&&{}
-rb_1^+
    \sum_{k=0}^{\infty}
    \left(\frac{-8r}{m_1^2}\right)^{k}
    \frac{(b_2^+)^k\;b_1^{2k+1}}{(2k+1)!}\;(2h-4^{-k})
+4r\; \frac{\tilde{\gamma}m_0}{m_1}\;f\;
    \sum_{k=0}^{\infty}
    \left(\frac{-2r}{m_1^2}\right)^{k}
    \frac{(b_2^+)^{k+1}\;b_1^{2k+1}}{(2k+1)!}\nonumber
    \\
    &&{} + {r\left(h-\frac{1}{2}\right)}
    \sum_{k=0}^{\infty}
    \left(\frac{-2r}{m_1^2}\right)^{k+1}
    \frac{(b_2^+)^{k+1}\;b_1^{2k+2}}{(2k+2)!}
-2r\;(b_1^+)^2\sum_{k=0}^{\infty}
    \left(\frac{-8r}{m_1^2}\right)^k
    \frac{(b_2^+)^k\;b_1^{2k+2}}{(2k+2)!}
 \nonumber
\end{eqnarray}
\vspace{-2ex}
\begin{eqnarray}
\phantom{l_0'}
 &&{} -2rf^+f\;
    \sum_{k=0}^{\infty}
    \left(\frac{-2r}{m_1^2}\right)^{k}
    \left\{\frac{(h-\textstyle\frac{1}{2})}{(2k)!}
    +\frac{b_1^+b_1}{(2k+1)!}\right\}(b_2^+)^{k}b_1^{2k}
\nonumber
\\
&&{} +   \frac{m_0^2-r(h^2-\frac{1}{4})}{2}
    \sum_{k=0}^{\infty}
    \left(\frac{-8r}{m_1^2}\right)^{k+1}
    \frac{(b_2^+)^{k+1}\;b_1^{2k+2}}{(2k+2)!}
,
\label{l0'}
\end{eqnarray}
\vspace{-2ex}
\begin{eqnarray}
l_1'
&=&
-m_1b_1^+b_2
+\frac{m_1}{4}\;b_1^+\;
    \sum_{k=1}^{\infty}
    \left(\frac{-8r}{m_1^2}\right)^{k}
    \left\{\frac{2h-4^{-k}}{(2k)!}
   +\frac{2b_1^+b_1}{(2k+1)!}\right\}(b_2^+)^{k-1}b_1^{2k}
+
\nonumber
\\
&&{}
+\frac{\tilde{\gamma}m_0}{4}\;(f^+{-}2b_2^+f)
    \sum_{k=1}^{\infty}
    \left(\frac{-8r}{m_1^2}\right)^{k}
    \frac{(b_2^+)^{k-1}\;b_1^{2k}}{(2k)!}\;(1-4^{-k})
\nonumber
\\
&&{}
+ \frac{r(h-\frac{1}{2})}{2m_1}
    \sum_{k=0}^{\infty}
    \left(\frac{-2r}{m_1^2}\right)^{k}
    \frac{(b_2^+)^{k}\;b_1^{2k+1}}{(2k+1)!}
+\frac{m_1}{2}\;b_1^+f^+f\;
     \sum_{k=1}^{\infty}
    \left(\frac{-2r}{m_1^2}\right)^{k}
    \frac{(b_2^+)^{k-1}b_1^{2k}}{(2k)!}
\nonumber
\\
&&{}
-\frac{r(h-{\textstyle\frac{1}{2}})}{m_1}f^+f
    \sum_{k=0}^{\infty}
    \left(\frac{-2r}{m_1^2}\right)^{k}
    \frac{(b_2^+)^{k}b_1^{2k+1}}{(2k+1)!}
-\tilde{\gamma}m_0f\sum_{k=0}^{\infty}
    \left(\frac{-2r}{m_1^2}\right)^{k}
    \frac{(b_2^+)^{k}b_1^{2k}}{(2k)!}
\nonumber
\\
&&{} +\frac{m_0^2-r(h^2-\frac{1}{4})}{m_1}
    \sum_{k=0}^{\infty}
    \left(\frac{-8r}{m_1^2}\right)^{k}
    \frac{(b_2^+)^{k}\;b_1^{2k+1}}{(2k+1)!}
,
\label{l1'}
\end{eqnarray}
\vspace{-2ex}
\begin{eqnarray}
l_2' &=& g_0'b_2 - b_2^+b_2^2
-\frac{m_0^2-r(h^2-\frac{1}{4})}{m_1^2}\;
    \sum_{k=0}^{\infty}
    \left(\frac{-8r}{m_1^2}\right)^{k}
    \frac{(b_2^+)^k\;b_1^{2k+2}}{(2k+2)!}
\nonumber
\\
&&{}
-\frac{r(h-\frac{1}{2})}{2m_1^2}\;
    \sum_{k=0}^{\infty}
    \left(\frac{-2r}{m_1^2}\right)^{k}
    \frac{(b_2^+)^k\;b_1^{2k+2}}{(2k+2)!}
+\frac{\tilde{\gamma}m_0}{m_1}\;f\;\sum_{k=0}^{\infty}
    \left(\frac{-2r}{m_1^2}\right)^{k}
    \frac{(b_2^+)^{k}b_1^{2k+1}}{(2k+1)!}
\nonumber
\\
&&{}
-  \frac{\tilde{\gamma}m_0}{4m_1}\;(f^+{-}2b_2^+f)
    \sum_{k=1}^{\infty}
    \left(\frac{-8r}{m_1^2}\right)^{k}
    \frac{(b_2^+)^{k-1}b_1^{2k+1}}{(2k+1)!}\;(1- 4^{-k})
\nonumber
\\
&&{}
-\frac{1}{4}\;b_1^+\;
    \sum_{k=1}^{\infty}
    \left(\frac{-8r}{m_1^2}\right)^{k}
    \left\{\frac{2h-4^{-k}}{(2k+1)!}
    +\frac{2b_1^+b_1}{(2k+2)!}
    \right\}(b_2^+)^{k-1}b_1^{2k+1}
\nonumber
\\
&&{}
- \frac{1}{2}\;f^+f\; \sum_{k=1}^{\infty}
    \left(\frac{-2r}{m_1^2}\right)^{k}
    \left\{\frac{h-\textstyle\frac{1}{2}}{(2k)!}
  + \frac{b_1^+b_1}{(2k+1)!}\right\}
    (b_2^+)^{k-1}b_1^{2k}
 \label{l2'}.
\end{eqnarray}
In the above expressions, $h$ is an arbitrary dimensionless
constant, while $m_0$ and $m_1$ are arbitrary  constants with
the dimension of mass.
 To obtain explicit expressions for the additional
parts (\ref{t1'+})--(\ref{l2'}) in the Fock space, we have
introduced two new pairs of bosonic and a new pair of fermionic
creation and annihilation operators satisfying the standard
commutation relations
\begin{eqnarray}
[b_1,b_1^+]=1,
\qquad
[b_2,b_2^+]=1,
\qquad
\{f, f^+\}=1.
\end{eqnarray}
 The resulting additional
parts of the operators possess all the necessary properties, in
particular, the additional parts which correspond to Hermitian
operators contain arbitrary parameters (see the footnote
\thefootnote): the operator $t_0'$ contains the arbitrary
parameter $m_0$; the operator $g_0'$ contains the arbitrary
parameter $h$. The values of the arbitrary parameters $h$ and
$m_0$ will be determined later by the condition that the correct
equations of motion (\ref{Eq-2}), or, equivalently, (\ref{TheEq}),
be reproduced.

The massive parameter $m_1$ remains arbitrary and can be expressed
in terms of other parameters of the theory,
\begin{eqnarray}
m_1=f(m,r)\neq0.
\label{m1}
\end{eqnarray}
This arbitrariness does not affect  the equations  for the basic
vector (\ref{PhysState}).

Note that the additional parts do not obey the usual properties
\begin{align}
&
(l_0')^+\neq l_0',
&&
(l_1')^+\neq l_1^{\prime+},
&&
(l_2')^+\neq l_2^{\prime+},
\\
&
(t_0')^+\neq t_0'
&&
(t_1')^+\neq t_1^{\prime+},
\end{align}
if one should use the standard rules of Hermitian conjugation for the new
creation and annihilation operators,
\begin{equation}
(b_1)^+=b_1^+,
\qquad
(b_2)^+=b_2^+,
\qquad
(f)^+=f^+.
\end{equation}
To restore the proper Hermitian conjugation properties for the
additional parts, we change the scalar product in the Fock space
$\mathcal{H}'$  generated by the new creation and annihilation
operators  as follows:
\begin{eqnarray}
\langle\tilde{\Psi}_1|\Psi_2\rangle_{\mathrm{new}} =
\langle\tilde{\Psi}_1|K'|\Psi_2\rangle\,, \label{newsprod}
\end{eqnarray}
for any vectors $|\Psi_1\rangle, |\Psi_2\rangle$ with some, yet
unknown, operator $K'$. This operator is determined by the
condition that all of the operators of the algebra  must have the
proper Hermitian properties with respect to the new scalar product:
\begin{align}
& \langle\tilde{\Psi}_1|K'l_0'|\Psi_2\rangle =
\langle\tilde{\Psi}_2|K'l_0'|\Psi_1\rangle^* , &&
\langle\tilde{\Psi}_1|K't_0'|\Psi_2\rangle =
\langle\tilde{\Psi}_2|K't_0'|\Psi_1\rangle^* ,
\\
& \langle\tilde{\Psi}_1|K'l_1'|\Psi_2\rangle =
\langle\tilde{\Psi}_2|K'l_1^{\prime+}|\Psi_1\rangle^* , &&
\langle\tilde{\Psi}_1|K't_1'|\Psi_2\rangle =
\langle\tilde{\Psi}_2|K't_1^{\prime+}|\Psi_1\rangle^* ,
\\
& \langle\tilde{\Psi}_1|K'l_2'|\Psi_2\rangle =
\langle\tilde{\Psi}_2|K'l_2^{\prime+}|\Psi_1\rangle^* , &&
\langle\tilde{\Psi}_1|K'g_0'|\Psi_2\rangle =
\langle\tilde{\Psi}_2|K'g_0'|\Psi_1\rangle^*
\end{align}
These relations permit one to determine the operator $K'$ as follows:
\begin{eqnarray}
\label{explicit K} K'=Z^+Z, \qquad
Z=\sum_{n_1,n_2=0}^{\infty}\sum_{s=0}^1
\left(\frac{{l^{\prime+}_1}}{m_1} \right)^{n_1}(t_1^{\prime+})^{s}
(l_2^{\prime+})^{n_2}|0\rangle_V\frac{1}{n_1!n_2!}\langle
0|b_1^{n_1}f^sb_2^{n_2},\;
\end{eqnarray}
where the auxiliary vector $|0\rangle_V$ obeys the relations
\begin{eqnarray}
&& t'_1|0\rangle_V = l_1'|0\rangle_V = l_2'|0\rangle_V = 0,
\qquad {}_V\langle0| 0\rangle_V =1
\\
&& {t}_0'|0\rangle_V=\tilde{\gamma}m_0|0\rangle_V,
\qquad
l_0'|0\rangle_V=m_0^2|0\rangle_V,
\qquad
g_0'|0\rangle_V=h|0\rangle_V
.
\label{0V}
\end{eqnarray}
For low numbers $n_1+n_f+2n_2$, where $n_i$ are the numbers of
``particles'' associated with $b_i^+$, and $n_f$ is the number of
``particles'' associated with $f^+$, the operator $K'$ reads
\begin{eqnarray}
K' &=& |0\rangle\langle0| +\frac{m_0^2-rh(h-\frac{1}{2})}{m_1^2}\;
  b_1^+|0\rangle\langle0|b_1
-2h f^+|0\rangle\langle  0|f
\nonumber
 \\
&&{}  +\frac{m_0}{m_1}
 \Bigl(\  \tilde{\gamma}f^+|0\rangle\langle0|b_1
  +b_1^+|0\rangle\langle0|f\tilde{\gamma}
  \Bigr)
+ \ldots\,.
\label{K}
\end{eqnarray}
This expression for the operator $K'$ will be used later in
constructing of examples in section~\ref{examples}.

Thus, in this section, we have constructed  the additional parts
(\ref{t1'+})--(\ref{l2'}) for the constraints, which obey all
the requirements. In the next section, we determine the algebra
of the enlarged constraints and construct a BRST operator
corresponding to this algebra.

\section{Deformed algebra and BRST operator}\label{BRST}
Let us turn to the algebra of the enlarged operators
$\tilde{O}_i=\tilde{o}_i+o_i'$. Since the algebra is quadratic,
there exist different possibilities of operator ordering in
the right-hand side of the commutation relations.
Analogous situation takes place in the bosonic higher spin
theory as well.
In \cite{0608005} we have studied dependence of BRST
construction on ordering prescription for the quadratic
constraint algebra and proved that different ordering
prescriptions lead to equivalent Lagrangians and the same
solution to the equations of motion.
Since the constraint algebra in the fermionic case is analogous
to the bosonic one, the same consequence about the final 
Lagrangians and the equations of motion will be valid also for
the fermionic case.
Therefore we are not going to study here all the possibilities
of ordering prescriptions as this was done in \cite{0608005},
but choose only one of them, which
corresponds to the supersymmetrized ordering for the constraints and
lead to the expression for the BRST operator which has
nonvanishing terms of the third degree in powers of 
ghosts\footnote{Using our
results in the bosonic case \cite{0608005}, we see that the
BRST operators corresponding to different choices of the mentioned
ordering of constraints will lead to equivalent Lagrangian
formulations. At the same time, we note that for the purpose of
constructing Lagrangians it is sufficient to find only one BRST
operator with a some fixed ordering for the constraints in its
commutation relations.}. 
The algebra of the enlarged operators
corresponding to this
 BRST operator is given by Table~\ref{Table},
\begin{table}
\begin{eqnarray*}
&
\begin{array}{||c||r|r|r|r|r|r|r|r||c||}\hline\hline\vphantom{\biggm|}
       [\;\downarrow\;,\to\}&T_0&T_1&
       T^+_1& L_0&L_1&L^+_1&L_2& L^+_2 &G_0\\
\hline\hline\vphantom{\biggm|}
  T_0
          &-2L_0&2L_1&2L_1^+&0&(\ref{T0L1})&
          - (\ref{L1+T0})
          & 0 & 0 &0\\
\hline\vphantom{\biggm|} T_1
                    &2L_1&4L_2&-2G_0& (\ref{T1L0})&0&
                    -T_0 & 0 & -T^+_1     & T_1 \\
\hline\vphantom{\biggm|} T^{+}_1
          &2{L}_1^+&-2G_0&4L^{+}_2&-(\ref{L0T1+})&T_0& 0 &
          T_1 & 0  &-T^{+}_1 \\
\hline\vphantom{\biggm|}
   L_{0}
        &0&-(\ref{T1L0})&(\ref{L0T1+})&0&-(\ref{L1L0})&(\ref{L0L1+})
        & 0 & 0 & 0 \\
\hline\vphantom{\biggm|}
   L_1
   &-(\ref{T0L1})& 0 &
   -T_0&(\ref{L1L0})&0&(\ref{L1L1+}) & 0 &
-L^{+}_1 &L_1 \\
\hline\vphantom{\biggm|}
  L^+_{1} &
    (\ref{L1+T0})&T_0& 0
&-(\ref{L0L1+})&-(\ref{L1L1+})& 0 & L_1 &0&
-L^{+}_1  \\
\hline\vphantom{\biggm|}
   L_{2}
&0&0&-T_1& 0 & 0 & -L_1 & 0 & G_0 &2L_2 \\
\hline\vphantom{\biggm|}
   L^{+}_{2}
   & 0 & T^{+}_1 & 0 & 0 & L^{+}_1
& 0 & -G_0 &0&
-2L^{+}_2 \\
\hline\hline\vphantom{\biggm|}
   G_{0}
&0&-T_1&T^{+}_1&0&-L_1&L^{+}_1 & -2L_2 & 2L^{+}_2&0\\
\hline\hline
\end{array}
\end{eqnarray*}
\caption{The algebra of the enlarged operators.} \label{Table}
\end{table}
where
\begin{eqnarray}
[T_0,L_1\} &=& r\Bigl[ \textstyle\frac{1}{2}G_0T_1
+\textstyle\frac{1}{2}T_1G_0 +T^{ +}_1L_2 +L_2T^{ +}_1 -g_0'T_1 -
t_1'G_0 - 2(t^{\prime+}_1L_2 + l'_2T^{ +}_1) \Bigr] , \label{T0L1}
\\
{} [L^+_1,T_0\} &=& r\Bigl[ \textstyle\frac{1}{2}T^{ +}_1G_0
+\textstyle\frac{1}{2}G_0T^{+}_1 +L^+_2T_1+T_1L^+_2 -t^{\prime
+}_1G_0 -g'_0T^+_1-2(l^{\prime+}_2T_1+t'_1L^+_2) \Bigr] ,
\label{L1+T0}
\\
{} [T_1,L_0\} &=& r\Bigl[ G_0T_1+T_1G_0+2T^+_1L_2+2L_2T^+_1
 -2\left(g'_0T_1 + t'_1G_0\right)
 -4(t^{\prime +}_1L_2 + l'_2T^+_1)
\Bigr]
,
\label{T1L0}
\\
{} [L_0,T^+_1\} &=& r\Bigl[\hspace{-0.1em} T^+_1G_0+G_0T^{
+}_1+2L^+_2T_1 +2T_1L^{ +}_2 -2 (t^{\prime+}_1G_0 + g'_0T^{ +}_1)
-4(l^{\prime +}_2T_1+t'_1L^{ +}_2)\hspace{-0.1em} \Bigr],
\label{L0T1+}
\\
{} [L_1,L_0\} &=& r\Bigl[ G_0L_1+L_1G_0 + 2L^+_1L_2 + 2 L_2L^{+}_1
-2\left(l'_1G_0+ g'_0L_1\right) -4\left(l^{\prime+}_1L_2 +
l'_2L^{+}_1\right) \Bigr] \hspace{-0.1em}, \label{L1L0}
\end{eqnarray}
\vspace{-3ex}
\begin{eqnarray}
{} [L_0,L^+_1\} &=& r\Bigl[ \hspace{-0.2em}L^+_1G_0+G_0L^+_1 +2L^{
+}_2L_1 + 2L_1L^+_2 -2\hspace{-0.2em}\left(l^{\prime
+}_1G_0+g'_0L^+_1\right)\hspace{-0.1em}
-4\hspace{-0.1em}\left(l^{\prime }_1L^{+}_2 +
l^{\prime+}_2L_1\right) \hspace{-0.2em}\Bigr]\hspace{-0.1em},
\label{L0L1+}
\\
 {} [L_1,L^+_1\} &=& L_0 + r \Bigl\{ G^{2}_0
-2L^+_2L_2 -2L_2L^{ +}_2 +{\textstyle\frac{3}{4}}T^{+}_1T_1
-{\textstyle\frac{3}{4}}T_1T^{+}_1 \Bigr\} \nonumber
\\
&&
\quad{}+
r
\Bigl\{
-2g'_0G_0 +4\left(l^{\prime +}_2L_2 + l'_2L^{ +}_2\right)
-{\textstyle\frac{3}{2}}\left(t^{\prime+}_1T_1 - t'_1T^{+}_1\right)
\Bigr\}
.
\label{L1L1+}
\end{eqnarray}

The construction of a nilpotent fermionic BRST operator for a
nonlinear  superalgebra
is based on the same principles as those developed in
\cite{B-BRST-Ads, 0608005} (for a general consideration of
operator BFV quantization, see the reviews \cite{bf}). The BRST operator
constructed on a basis of the algebra given by
Table~\ref{Table} can be calculated with the help of the $(\mathcal{C}
\mathcal{P})$-ordering of the ghost coordinate $\mathcal{C}^i$
and momenta $\mathcal{P}_i$ operators:
\begin{eqnarray}
\label{Q'}
Q' \hspace{-0.4em} &=&\hspace{-0.4em}
q_0T_0+q_1^+T_1+q_1T_1^+ +\eta_0L_0+\eta_1^+L_1+\eta_1L_1^+
+\eta_2^+L_2 +\eta_2L_2^+ +\eta_{G}G_0 \nonumber
\\\hspace{-0.4em}&&
{}\hspace{-0.4em} +i(\eta_1^+q_1-\eta_1q_1^+)p_0
-i(\eta_Gq_1+\eta_2q_1^+)p_1^+ +i(\eta_Gq_1^++\eta_2^+q_1)p_1
\nonumber
\\
\hspace{-0.4em}&&{}\hspace{-0.4em}
+(q_0^2-\eta_1^+\eta_1){\cal{}P}_0
+(2q_1q_1^+-\eta_2^+\eta_2){\cal{}P}_G
+(\eta_G\eta_1^++\eta_2^+\eta_1-2q_0q_1^+){\cal{}P}_1 \nonumber
\\
\hspace{-0.4em}&&\hspace{-0.4em}
+(\eta_1\eta_G+\eta_1^+\eta_2-2q_0q_1){\cal{}P}_1^+
+2(\eta_G\eta_2^+-q_1^{+2}){\cal{}P}_2
+2(\eta_2\eta_G-q_1^2){\cal{}P}_2^+ \nonumber\\
\hspace{-0.4em} &+& \hspace{-0.4em}r(q_0 \eta_1^+ + 2q_1^+ \eta_0)
    \hspace{-0.1em}\Bigl[
     {\textstyle\frac{i}{2}}\,G_0p_1
    +{\textstyle\frac{1}{2}}\,T_1\mathcal{P}_G
    +T_1^+\mathcal{P}_2
    +iL_2p_1^+
    -t_1'\mathcal{P}_G
    -ig_0'p_1
    -2(il_2'p_1^+ +t_1^{\prime+}\mathcal{P}_2 )
\hspace{-0.1em}\Bigr] \nonumber
\\
\hspace{-0.4em}&+& \hspace{-0.4em} r(q_0 \eta_1+ 2q_1 \eta_0 )
\hspace{-0.1em}\Bigl[
    -{\textstyle\frac{i}{2}}G_0p_1^+
    -{\textstyle\frac{1}{2}}T_1^+\mathcal{P}_G
    -T_1\mathcal{P}_2^+
    -iL_2^+p_1
    +t_1^{\prime +}\mathcal{P}_G
    +ig_0'p_1^+
    +2 (il_2^{\prime +}p_1+t_1' \mathcal{P}_2^+ )
\hspace{-0.1em}\Bigr] \nonumber
\\
\hspace{-0.4em}&&{}\hspace{-0.4em} +2r\eta_0 \eta_1^+ \Bigl[
     {\textstyle\frac{1}{2}}G_0\mathcal{P}_1
     +{\textstyle\frac{1}{2}}L_1\mathcal{P}_G
     +L_1^+\mathcal{P}_2
     +L_2\mathcal{P}_1^+
     -l_1'\mathcal{P}_G
     -g_0'\mathcal{P}_1
     -2 (l_1^{\prime +}\mathcal{P}_2
     +l_2'\mathcal{P}_1^+ )
\Bigr]
\nonumber
\\
\hspace{-0.4em}&&{}\hspace{-0.4em} -2r\eta_0\eta_1 \Bigl[
     {\textstyle\frac{1}{2}}G_0\mathcal{P}_1^+
    +{\textstyle\frac{1}{2}}L_1^+\mathcal{P}_G
    +L_1\mathcal{P}_2^+
    +L_2^+\mathcal{P}_1
    - l_1^{\prime +}\mathcal{P}_G
    -g_0'\mathcal{P}_1^+
    -2(l_2^{\prime +}\mathcal{P}_1
    +l_1' \mathcal{P}_2^+ )
\Bigr]
\nonumber
\\
\hspace{-0.4em}&&{} \hspace{-0.4em} -r\eta_1\eta_1^+ \Bigl[
     2 L_2^+ \mathcal{P}_2
   + 2 L_2\mathcal{P}_2^+
   - G_0\mathcal{P}_G
   - {\textstyle\frac{3i}{4}}T_1^+p_1
   + {\textstyle\frac{3i}{4}}T_1p_1^+
\Bigr]
\nonumber
\\
\hspace{-0.4em}&&{}\hspace{-0.4em} +r\eta_1\eta_1^+ \Bigl[
      4( l_2^{\prime+}\mathcal{P}_2+l'_2\mathcal{P}_2^+)
    - 2 g'_0\mathcal{P}_G
    - {\textstyle\frac{3i}{2}}(t_1^{\prime +}p_1 - t_1'p_1^+)
\Bigr]
\nonumber
\\
\hspace{-0.4em}&+& \hspace{-0.4em}r^2 \eta_0\eta_1\eta^+_1
  \left[G_0
   \left(p_1^+p_1+2\mathcal{P}^+_2\mathcal{P}_2\right)
   + L_2^+p_1^2
   + L_2(p_1^+)^2
   +\frac{i}{2}\left(T_1^+p_1^+\mathcal{P}_2+T_1p_1\mathcal{P}^+_2\right)
\right.
\nonumber
\\
\hspace{-0.4em}&&{}\hspace{-0.4em}
 \qquad\qquad\left.
   +\frac{i}{4}\left(T_1^+p_1 + T_1p^+_1\right)\mathcal{P}_G
   -2\left(L_2\mathcal{P}_2^+-L_2^+\mathcal{P}_2\right)\mathcal{P}_G
  \right]
.
\end{eqnarray}
Here, $q_0$, $q_1$, $q_1^+$ and  $\eta_0$, $\eta_1^+$, $\eta_1$,
$\eta_2^+$, $\eta_2$, $\eta_G$ are, respectively, the bosonic and
fermionic ghost ``coordinates'' corresponding to their canonically
conjugate ghost ``momenta'' $p_0$, $p_1^+$, $p_1$, ${\cal{}P}_0$,
${\cal{}P}_1$, ${\cal{}P}_1^+$, ${\cal{}P}_2$, ${\cal{}P}_2^+$,
${\cal{}P}_G$. They obey the (anti)commutation relations
\begin{eqnarray}
\label{ghosts}
&
\{\eta_1,{\cal{}P}_1^+\}= \{{\cal{}P}_1, \eta_1^+\}
=
\{\eta_2,{\cal{}P}_2^+\}= \{{\cal{}P}_2, \eta_2^+\}
=\{\eta_0,{\cal{}P}_0\}= \{\eta_G,{\cal{}P}_G\} =
1,
\nonumber
\\
& [q_0, p_0]=[q_1, p_1^+] = [q_1^+, p_1] = i
\end{eqnarray}
and possess the standard  ghost number distribution,
$gh(\mathcal{C}^i)$ = $ - gh(\mathcal{P}_i)$ = $1$,
providing the property  $gh(\tilde{Q}')$ = $1$.

The  resulting BRST operator $Q'$ is Hermitian with respect to the
scalar product $\langle \ |\ \rangle$  in the tensor product of
the Fock spaces $\mathcal{H}_{tot} = \mathcal{H}'\otimes
\mathcal{H}\otimes \mathcal{H}_{gh}$,  which  is constructed as
the direct product of the scalar products on $\mathcal{H},
\mathcal{H}'$ and $\mathcal{H}_{gh}$,
\begin{eqnarray}
\langle\tilde{\Psi}_1| Q^{\prime +}K|\Psi_2\rangle =
\langle\tilde{\Psi}_1|K Q'|\Psi_2\rangle
\end{eqnarray}
with the operator $K$ defined in $\mathcal{H}_{tot}$ and being
the tensor product of the operator $K'$ and the unit operators $\hat{1},
\hat{1}_{gh}$,
\begin{eqnarray} \label{tK}
  K &=&  K' \otimes \hat{1} \otimes \hat{1}_{gh}.
\end{eqnarray}
Thus, we have constructed the Hermitian BRST operator. In the next
section, this operator is applied to construct Lagrangians of
fermionic higher spin fields in the AdS space.

\section{Construction of Lagrangians}\label{Lagr-constr}

In this section, we construct Lagrangians of fermionic massive
higher spin fields in the AdS space.
This construction goes along the lines of
\cite{0410215,0603212}.
We should first extract the dependence of the BRST
operator $Q'$ (\ref{Q'}) on the ghosts $\eta_G$, $\mathcal{P}_G$,
\begin{eqnarray}
Q'&=&Q+\eta_G(\sigma+h)+\mathcal{B}\mathcal{P}_G,
\end{eqnarray}
where
\begin{eqnarray}
\label{Q} Q &=& q_0T_0+q_1^+T_1+q_1T_1^+
+\eta_0L_0+\eta_1^+L_1+\eta_1L_1^+ +\eta_2^+L_2 +\eta_2L_2^+
\nonumber
\\&&
{}
+i(\eta_1^+q_1-\eta_1q_1^+)p_0
-i\eta_2q_1^+p_1^+
+i\eta_2^+q_1p_1
+(q_0^2-\eta_1^+\eta_1){\cal{}P}_0
+(\eta_2^+\eta_1-2q_0q_1^+){\cal{}P}_1
\nonumber
\\
&&{} +(\eta_1^+\eta_2-2q_0q_1){\cal{}P}_1^+ -2q_1^{+2}{\cal{}P}_2
-2q_1^2{\cal{}P}_2^+ \nonumber \\
&+& r(q_0 \eta_1^+ + 2q_1^+
\eta_0)
    \Bigl[
     {\textstyle\frac{i}{2}}\,G_0p_1
    +T_1^+\mathcal{P}_2
    +iL_2p_1^+
    -ig_0'p_1
    -2(il_2'p_1^+ +t_1^{\prime+}\mathcal{P}_2 )
\Bigr]
\nonumber
\\
&+&
r(q_0 \eta_1+ 2q_1 \eta_0 )
\Bigl[
    -{\textstyle\frac{i}{2}}G_0p_1^+
    -T_1\mathcal{P}_2^+
    -iL_2^+p_1
    +ig_0'p_1^+
    +2 (il_2^{\prime +}p_1+t_1' \mathcal{P}_2^+ )
\Bigr] \nonumber\\
&&{} +2r\eta_0 \eta_1^+ \Bigl[
     {\textstyle\frac{1}{2}}G_0\mathcal{P}_1
     +L_1^+\mathcal{P}_2
     +L_2\mathcal{P}_1^+
     -g_0'\mathcal{P}_1
     -2 (l_1^{\prime +}\mathcal{P}_2
     +l_2'\mathcal{P}_1^+ )
\Bigr] \nonumber \\
 &&{} -2r\eta_0\eta_1 \Bigl[
     {\textstyle\frac{1}{2}}G_0\mathcal{P}_1^+
    +L_1\mathcal{P}_2^+
    +L_2^+\mathcal{P}_1
    -g_0'\mathcal{P}_1^+
    -2(l_2^{\prime +}\mathcal{P}_1
    +l_1' \mathcal{P}_2^+ )
\Bigr]
\nonumber
\end{eqnarray}
\vspace{-3ex}
\begin{eqnarray}
\phantom{Q} &&{} -r\eta_1\eta_1^+ \Bigl[
     2 L_2^+ \mathcal{P}_2
   + 2 L_2\mathcal{P}_2^+
   - {\textstyle\frac{3i}{4}}T_1^+p_1
   + {\textstyle\frac{3i}{4}}T_1p_1^+
\Bigr]
\nonumber
\\
&&{}
+r\eta_1\eta_1^+
\Bigl[
      4( l_2^{\prime+}\mathcal{P}_2+l'_2\mathcal{P}_2^+)
    - {\textstyle\frac{3i}{2}}(t_1^{\prime +}p_1 - t_1'p_1^+)
\Bigr]
\nonumber
\\
&+&
r^2
\eta_0\eta_1\eta^+_1
  \left[G_0
   \left(p_1^+p_1+2\mathcal{P}^+_2\mathcal{P}_2\right)
   + L_2^+p_1^2
   + L_2(p_1^+)^2
   +\frac{i}{2}\left(T_1^+p_1^+\mathcal{P}_2+T_1p_1\mathcal{P}^+_2\right)
  \right],
\end{eqnarray}
\vspace{-4ex}
\begin{eqnarray}
\label{sigma}
\sigma
&=&
-a^+_\mu a^\mu +\frac{d}{2} +b_1^+b_1+2b_2^+b_2+f^+f
\nonumber
\\
&&{}
-iq_1p_1^++iq_1^+p_1
+\eta_1^+\mathcal{P}_1-\eta_1\mathcal{P}_1^+
+2\eta_2^+\mathcal{P}_2-2\eta_2\mathcal{P}_2^+\,;
\end{eqnarray}
meanwhile the explicit expression for the operator $\mathcal{B}$ is not
essential.

Next, following the procedure of \cite{0410215,
0603212}, we choose the following representation of the Hilbert space:
\begin{equation}
\left( p_0, q_1, p_1, \mathcal{P}_0,
{\cal{}P}_G, \eta_1, {\cal{}P}_1, \eta_2,
{\cal{}P}_2\right)|0\rangle =0\,,
\label{ghostvac}
\end{equation}
and suppose
that the  vectors and gauge parameters do not depend on  $\eta_G$,
\begin{eqnarray} \label{chi}
&  |\chi\rangle = \displaystyle\sum\limits_{k_i}
(q_0)^{k_1}(q_1^+)^{k_2}(p_1^+)^{k_3}(\eta_0)^{k_4}(f^+)^{k_5}
  (\eta_1^+)^{k_6}(\mathcal{P}_1^+)^{k_7}(\eta_2^+)^{k_8} (\mathcal{P}_2^+)^{k_9}
  (b_1^+)^{k_{10}}(b_2^+)^{k_{11}} \times&\nonumber \\
   & \times a^{+{}\mu_1}\cdots a^{+{}\mu_{k_0}}\chi^{k_1 \cdots k_{11}}_{\mu_1\cdots \mu_{k_0}}(x)
   |0\rangle. &
\end{eqnarray}
The sum in (\ref{chi}) is taken  over $k_0, k_1, k_2$, $k_3$,
$k_{10}$, $k_{11}$, running from 0 to infinity, and over $k_4, k_5,
k_6, k_7, k_8, k_9$, running from 0 to 1.
Then, we derive from the
equations that determine the physical  vector, $Q'|\chi\rangle$ =
$0$, as well as from the reducible gauge transformations, $\delta|\chi\rangle$
= $Q'|\Lambda\rangle$, a sequence of relations:
\begin{align}
\label{Qchi} & Q|\chi\rangle=0, && (\sigma+h)|\chi\rangle=0, &&
\left(\varepsilon, {gh}\right)(|\chi\rangle)=(1,0),
\\
& \delta|\chi\rangle=Q|\Lambda\rangle, &&
(\sigma+h)|\Lambda\rangle=0, && \left(\varepsilon,
{gh}\right)(|\Lambda\rangle)=(0,-1), \label{QLambda}
\\
& \delta|\Lambda\rangle=Q|\Lambda^{(1)}\rangle, &&
(\sigma+h)|\Lambda^{(1)}\rangle=0, && \left(\varepsilon,
{gh}\right)(|\Lambda^{(1)}\rangle)=(1,-2),\\
& \delta|\Lambda^{(i-1)}\rangle=Q|\Lambda^{(i)}\rangle, &&
(\sigma+h)|\Lambda^{(i)}\rangle=0, && \left(\varepsilon,
{gh}\right)(|\Lambda^{(i)}\rangle)= (i,-i-1). \label{QLambdai}
\end{align}
The middle equation in (\ref{Qchi})
presents the equations for the possible values of $h$,
\begin{eqnarray}
\label{h}
-h&=&n+\frac{d-4}{2}
\;,
\end{eqnarray}
with $n$ being related to spin, $s=n+1/2$.
By fixing the value of spin, we also fix the parameter $h$, according to
(\ref{h}).
Having fixed a value of $h$, we must substitute it into each of
the expressions (\ref{Qchi})--(\ref{QLambdai});
see \cite{0410215} for more details.

As a next step, we have to extract the zero-mode ghosts from the
operator $Q$ (\ref{Q}). This operator has the structure
\begin{eqnarray}
\label{strQ}
Q&=&q_0\tilde{T}_0+\eta_0\tilde{L}_0
+i(\eta_1^+q_1-\eta_1q_1^+)p_0 +(q_0^2-\eta_1^+\eta_1){\cal{}P}_0
+\Delta{}Q,
\end{eqnarray}
where
\begin{eqnarray}
\label{tildeT0}
\tilde{T}_0 &=& T_0 -2q_1^+{\cal{}P}_1 -2q_1{\cal{}P}_1^+
\nonumber
\\
&&{}
+
r \eta_1^+
    \Bigl[
     {\textstyle\frac{i}{2}}\,G_0p_1
    +T_1^+\mathcal{P}_2
    +iL_2p_1^+
    -ig_0'p_1
    -2(il_2'p_1^+ +t_1^{\prime+}\mathcal{P}_2 )
\Bigr]
\nonumber
\\
&&{}
+
r \eta_1
\Bigl[
    -{\textstyle\frac{i}{2}}G_0p_1^+
    -T_1\mathcal{P}_2^+
    -iL_2^+p_1
    +ig_0'p_1^+
    +2 (il_2^{\prime +}p_1+t_1' \mathcal{P}_2^+ )
\Bigr],\phantom{T_1^+p_1^+\mathcal{P}_2+T_1p_1\mathcal{P}^+_2}
\end{eqnarray}
\vspace{-3ex}
\begin{eqnarray}
 \label{tildeL0} \tilde{L}_0 &=& L_0 + 2rq_1^+
    \Bigl[
     {\textstyle\frac{i}{2}}\,G_0p_1
    +T_1^+\mathcal{P}_2
    +iL_2p_1^+
    -ig_0'p_1
    -2(il_2'p_1^+ +t_1^{\prime+}\mathcal{P}_2 )
\Bigr]
\nonumber
\\
&&{}
+
2rq_1
\Bigl[
    -{\textstyle\frac{i}{2}}G_0p_1^+
    -T_1\mathcal{P}_2^+
    -iL_2^+p_1
    +ig_0'p_1^+
    +2 (il_2^{\prime +}p_1+t_1' \mathcal{P}_2^+ )
\Bigr]
\nonumber
\\
&&{}
+2r \eta_1^+
\Bigl[
     {\textstyle\frac{1}{2}}G_0\mathcal{P}_1
     +L_1^+\mathcal{P}_2
     +L_2\mathcal{P}_1^+
     -g_0'\mathcal{P}_1
     -2 (l_1^{\prime +}\mathcal{P}_2
     +l_2'\mathcal{P}_1^+ )
\Bigr]
\nonumber
\\
&&{}
-2r\eta_1
\Bigl[
     {\textstyle\frac{1}{2}}G_0\mathcal{P}_1^+
    +L_1\mathcal{P}_2^+
    +L_2^+\mathcal{P}_1
    -g_0'\mathcal{P}_1^+
    -2(l_2^{\prime +}\mathcal{P}_1
    +l_1' \mathcal{P}_2^+ )
\Bigr]
\nonumber
\\
&&{}
+
r^2
\eta_1\eta^+_1
  \left[G_0
   \left(p_1^+p_1+2\mathcal{P}^+_2\mathcal{P}_2\right)
   + L_2^+p_1^2
   + L_2(p_1^+)^2
   +\frac{i}{2}\left(T_1^+p_1^+\mathcal{P}_2+T_1p_1\mathcal{P}^+_2\right)
  \right],
\end{eqnarray}
\begin{eqnarray}
\label{DQ}
\Delta{}Q
&=&
q_1^+T_1+q_1T_1^+
+\eta_1^+L_1+\eta_1L_1^+
+\eta_2^+L_2
+\eta_2L_2^+
-i\eta_2q_1^+p_1^+
+i\eta_2^+q_1p_1
\nonumber
\\
&&{}
+\eta_2^+\eta_1{\cal{}P}_1
+\eta_1^+\eta_2{\cal{}P}_1^+
-2q_1^{+2}{\cal{}P}_2
-2q_1^2{\cal{}P}_2^+
\nonumber
\\
&&{}
-r\eta_1\eta_1^+
\Bigl[
     2 L_2^+ \mathcal{P}_2
   + 2 L_2\mathcal{P}_2^+
   - {\textstyle\frac{3i}{4}}T_1^+p_1
   + {\textstyle\frac{3i}{4}}T_1p_1^+
\Bigr]
\nonumber
\\
&&{}
+r\eta_1\eta_1^+
\Bigl[
      4( l_2^{\prime+}\mathcal{P}_2+l'_2\mathcal{P}_2^+)
    - {\textstyle\frac{3i}{2}}(t_1^{\prime +}p_1 - t_1'p_1^+)
\Bigr].
\end{eqnarray}
Here, $\tilde{T}_0$, $\tilde{L}_0$, $\Delta{}Q$ are independent
of $q_0$, $p_0$, $\eta_0$, $\mathcal{P}_0$.
We also expand the state vector and gauge parameters in powers
of the zero-mode ghosts:
\begin{align}
\label{0chi}
|\chi\rangle
&=\sum_{k=0}^{\infty}q_0^k(
|\chi_0^k\rangle
+\eta_0|\chi_1^k\rangle),
&
&gh(|\chi^{k}_{m}\rangle)=-(m+k),
\\
\label{0L}
|\Lambda^{(i)}\rangle
&=\sum_{k=0}^{\infty}q_0^k(|\Lambda^{(i)}{}^k_0\rangle
+\eta_0|\Lambda^{(i)}{}^k_1\rangle),
&
&gh(|\Lambda^{(i)}{}^k_m\rangle)=-(i+k+m+1)
.
\end{align}
Following the procedure of \cite{0410215}, we get rid of
all the fields except two, $|\chi^0_0\rangle$, $|\chi^1_0\rangle$,
and, hence, relations (\ref{Qchi})--(\ref{QLambdai}), with
allowance for (\ref{strQ}), (\ref{0chi}), (\ref{0L}), yield
two independent equations for these fields:
\begin{eqnarray}
&&
\Delta{}Q|\chi^{0}_{0}\rangle
+\frac{1}{2}\bigl\{\tilde{T}_0,\eta_1^+\eta_1\bigr\}
|\chi^{1}_{0}\rangle
=0,
\label{EofM1all}
\\&&
\tilde{T}_0|\chi^{0}_{0}\rangle
+
\Delta{}Q|\chi^{1}_{0}\rangle
=0,
\label{EofM2all}
\end{eqnarray}
where $\{A,B\}=AB+BA$ for any quantities $A, B$.

Next, due to the fact that the  operators $Q$, $\tilde{T}_0$,
$\eta_1^+\eta_1$ commute with  $\sigma$, we derive from
(\ref{EofM1all}), (\ref{EofM2all}) the equations of
motion for the fields with a fixed value of  spin:
\begin{eqnarray}
&&
\Delta{}Q|\chi^{0}_{0}\rangle_n
+\frac{1}{2}\bigl\{\tilde{T}_0,\eta_1^+\eta_1\bigr\}
|\chi^{1}_{0}\rangle_n
=0,
\label{EofM1}
\\&&
\tilde{T}_0|\chi^{0}_{0}\rangle_n +
\Delta{}Q|\chi^{1}_{0}\rangle_n =0. \label{EofM2}
\end{eqnarray}
where the fields $|\chi_0^0\rangle$, $|\chi_0^1\rangle$
are assumed to obey the relations
\begin{eqnarray}\label{schi}
\sigma|\chi^0_0\rangle_n =\bigl(n+(d-4)/2\bigr)|\chi^0_0\rangle_n,
&\qquad& \sigma|\chi^1_0\rangle_n
=\bigl(n+(d-4)/2\bigr)|\chi^1_0\rangle_n
.
\end{eqnarray}

The field equations (\ref{EofM1}), (\ref{EofM2}) are Lagrangian
ones and can be deduced from the following action:\footnote{The
action is defined, as usual, up to an overall factor.}
\begin{eqnarray}
{\cal{}S}_n &=&
{}_n\langle\tilde{\chi}^{0}_{0}|K_n\tilde{T}_0|\chi^{0}_{0}\rangle_n
+ \frac{1}{2}\,{}_n\langle\tilde{\chi}^{1}_{0}|K_n\bigl\{
   \tilde{T}_0,\eta_1^+\eta_1\bigr\}|\chi^{1}_{0}\rangle_n
\nonumber
\\&&
+
{}_n\langle\tilde{\chi}^{0}_{0}|K_n\Delta{}Q|\chi^{1}_{0}\rangle_n
+
{}_n\langle\tilde{\chi}^{1}_{0}|K_n\Delta{}Q|\chi^{0}_{0}\rangle_n
, \label{L1}
\end{eqnarray}
where the standard scalar product for the creation and
annihilation operators is assumed, and the operator $K_n$ is the
operator $K$ (\ref{tK}), where the following substitution is made:
$h\to-\bigl(n+(d-4)/2\bigr)$.

The equations of motion (\ref{EofM1}), (\ref{EofM2}) and the
action (\ref{L1}) are invariant with respect to the gauge
transformations
\begin{eqnarray}
\delta|\chi^{0}_{0}\rangle_n
&=&
\Delta{}Q|\Lambda^{0}_{0}\rangle_n
 +
 \frac{1}{2}\bigl\{\tilde{T}_0,\eta_1^+\eta_1\bigr\}
 |\Lambda^{1}_{0}\rangle_n,
\label{GT1}
\\
\delta|\chi^{1}_{0}\rangle_n
&=&
\tilde{T}_0|\Lambda^{0}_{0}\rangle_n
 +\Delta{}Q|\Lambda^{1}_{0}\rangle_n
 ,
\label{GT2}
\end{eqnarray}
which are reducible, with the gauge parameters
$|\Lambda^{(i)}{}^{j}_{0}\rangle_n$, $j=0,1$
subject to the same conditions
as those for $|\chi^j_0\rangle_n$ in (\ref{schi}),
\begin{align}
\delta|\Lambda^{(i)}{}^{0}_{0}\rangle_n
&=
\Delta{}Q|\Lambda^{(i+1)}{}^{0}_{0}\rangle_n
 +
 \frac{1}{2}\bigl\{\tilde{T}_0,\eta_1^+\eta_1\bigr\}
 |\Lambda^{(i+1)}{}^{1}_{0}\rangle_n,
&
|\Lambda^{(0)}{}^0_0\rangle_n=|\Lambda^0_0\rangle_n,
\label{GTi1}
\\
\delta|\Lambda^{(i)}{}^{1}_{0}\rangle_n
&=
\tilde{T}_0|\Lambda^{(i+1)}{}^{0}_{0}\rangle_n
 +\Delta{}Q|\Lambda^{(i+1)}{}^{1}_{0}\rangle_n,
&
|\Lambda^{(0)}{}^1_0\rangle_n=|\Lambda^1_0\rangle_n,
\label{GTi2}
\end{align}
and with a finite number of reducibility stages at $i_{max}=n-1$
for spin $s=n+1/2$.

We now determine the value of the arbitrary parameter $m_0$, using
the condition that the equations (\ref{TheEq}) for the basic vector
$|\Phi\rangle$ (\ref{PhysState}) be reproduced. To this end, it is
necessary that conditions (\ref{TheEq}) be implied by Eqs.
(\ref{EofM1}), (\ref{EofM2}). Note that the general vector
$|\chi^0_0\rangle_n$ includes the basic vector $|\Phi\rangle$
(\ref{PhysState}), namely,
\begin{equation}\label{relation}
    |\chi^0_0\rangle_n = |\Phi\rangle_n + |\Phi_A\rangle_n, \qquad
\left.\phantom{\Bigl[}|\Phi_A\rangle_n\right|_{\mathcal{C}=\mathcal{P}=b^+_1=b^+_2=f^+=0}=0
.
\end{equation}
In the next section, we shall demonstrate that due to the gauge transformations
and a part of the equations of motion the vector $|\Phi_A\rangle_n$ can be
completely removed, so that the resulting equations of motion have the form
\begin{equation}
\label{result eqs}  T_0|\Phi\rangle_n  =
(t_0+\tilde{\gamma}m_0)|\Phi\rangle_n =0,\qquad
{t}_1|\Phi\rangle_n =0.
\end{equation}
The above relations permit one to determine the parameter $m_0$ in a
unique way, as follows:
\begin{equation}
\label{final m0} m_0 = m-r^{\frac{1}{2}}h= m +
r^{\frac{1}{2}}\bigl(n+(d-4)/2\bigr).
\end{equation}
It should be noted that $m_0$ of the present article is the AdS mass $m_D$ of
\cite{0609029} corresponding to ``the most-used definition of
mass'' (for more discussions of this point, see \cite{0609029}).

In the next section, we shall demonstrate that the action actually
reproduces the correct equations of motion (\ref{Eq-2}). Thus, we
have constructed Lagrangians for fermionic fields of any fixed
spin using the BRST approach\footnote{The construction of a
Lagrangian describing the propagation of all fermionic fields in
the AdS space simultaneously is analogous to the case of the flat
space \cite{0410215} and we do not consider it here. We only note
that the necessary condition for resolving this problem is to
replace in $Q^{\prime}$, $Q$, $K$ the parameter $-h$ by the
operator $\sigma$ in an appropriate way and discard the condition
(\ref{schi}) for the fields and gauge parameters.}.

\section{Reduction to the initial irreducible relations}\label{reduction}

Let us show that the equations of motion (\ref{Eq-0}),
(\ref{Eq-1}) [or equivalently (\ref{TheEq})] can be obtained from
the Lagrangian (\ref{L1}) after gauge-fixing and removing the
auxiliary fields by using a part of the equations of motion. Let
us start with gauge-fixing.

\subsection{Gauge-fixing}

Let us consider a field of spin $s=n+1/2$. Then we have a
reducible gauge theory with $n-1$ reducibility stages. Due to
restriction (\ref{schi}) and the ghost number restriction [see the
right-hand formulae in (\ref{0L})],  the lowest-stage gauge
parameters have the form
\begin{eqnarray}
\label{gauge param}
|\Lambda_0^{(n-1){}0}\rangle_n
&=&
(p^+_1)^{n-1}\left\{\mathcal{P}_1^+|\lambda \rangle_0 +
{p}_1^+|\lambda_1
\rangle_0\right\}, \\
|\Lambda_0^{(n-1){}1}\rangle_n
& \equiv &0,
\end{eqnarray}
with the subscripts of the state vectors being
associated with the eigenvalues of the corresponding state vectors
(\ref{schi}).
It can be verified directly that
one can eliminate
the dependence on the ghost $\mathcal{P}_2^+$ from the gauge
function $|\Lambda^{(n-2)}{}^0_0\rangle$ of the $(n-2)$-th stage
(the gauge function $|\Lambda^{(n-2)}{}^1_0\rangle$ has no
$\mathcal{P}_2^+$ dependence).
It is then possible to verify that one can remove the dependence of
$|\Lambda^{(n-3)}{}^0_0\rangle$, $|\Lambda^{(n-3)}{}^1_0\rangle$
on $\mathcal{P}_2^+$ with the
help of the remaining gauge parameters (which do not depend on
$\mathcal{P}_2^+$)
$|\Lambda^{(n-2)}{}^0_0\rangle$,
$|\Lambda^{(n-2)}{}^1_0\rangle$.

We now suppose that we have removed the dependence on
$\mathcal{P}_2^+$ from the gauge functions of the $(i+1)$-th stage
$|\Lambda^{(i+1)}{}^j_0\rangle$, $j=0,1$, i.e., we have
$\eta_2|\Lambda^{(i+1)}{}^j_0\rangle=0$.
Let us show that these restricted gauge functions can be used
to remove the dependence on $\mathcal{P}_2^+$ from the gauge functions
$|\Lambda^{(i)}{}^j_0\rangle$.
We introduce the following notation for the
gauge parameters, related to their decomposition in ghosts
$\mathcal{P}_1^+$, $\mathcal{P}_2^+$:
\begin{eqnarray}
|\Lambda^{(i)}{}^j_0\rangle &=& |\Lambda^{(i)}{}^j_{00}\rangle
+\mathcal{P}_1^+|\Lambda^{(i)}{}^j_{01}\rangle
+\mathcal{P}_2^+|\Lambda^{(i)}{}^j_{02}\rangle
+\mathcal{P}_1^+\mathcal{P}_2^+|\Lambda^{(i)}{}^j_{03}\rangle.
\end{eqnarray}
Here and elsewhere, we omit the subscripts of the vectors which
are associated with the eigenvalues  of the operator $\sigma$
(\ref{schi}). Then, using (\ref{GTi1}) and (\ref{GTi2}), we find a
gauge transformation for $|\Lambda^{(i)}{}^j_{02}\rangle$ and
$|\Lambda^{(i)}{}^j_{03}\rangle$, being coefficients at
$\mathcal{P}_2^+$,
\begin{eqnarray}
\label{dL02}
\delta|\Lambda^{(i)}{}^0_{02}\rangle
&=&
-2q_1^2|\Lambda^{(i+1)}{}^0_{00}\rangle
+2r\eta_1^+(L_2-2l_2')|\Lambda^{(i+1)}{}^0_{01}\rangle
,
\\
\label{dL03}
\delta|\Lambda^{(i)}{}^0_{03}\rangle
&=&
2q_1^2|\Lambda^{(i+1)}{}^0_{01}\rangle
,
\\
\label{dL12}
\delta|\Lambda^{(i)}{}^1_{02}\rangle
&=&
-2q_1^2|\Lambda^{(i+1)}{}^1_{00}\rangle
+2r\eta_1^+(L_2-2l_2')|\Lambda^{(i+1)}{}^1_{01}\rangle
+r(T_1-2t_1')|\Lambda^{(i+1)}{}^0_{01}\rangle
\\
\label{dL13}
\delta|\Lambda^{(i)}{}^1_{03}\rangle
&=&
2q_1^2|\Lambda^{(i+1)}{}^1_{01}\rangle
.
\end{eqnarray}
Using (\ref{dL02})--(\ref{dL13}), we can see that the dependence on
$\mathcal{P}_2^+$ in $|\Lambda^{(i)}{}^j_0\rangle$ can be
removed with the help of the gauge transformations.
To this end, we should first make gauge transformations
with $|\Lambda^{(i+1)}{}^0_{01}\rangle$ and
$|\Lambda^{(i+1)}{}^1_{01}\rangle$, removing
$|\Lambda^{(i)}{}^0_{03}\rangle$ and
$|\Lambda^{(i)}{}^1_{03}\rangle$, respectively.
Then we should make gauge transformation with the parameter
$|\Lambda^{(i+1)}{}^0_{00}\rangle$, removing
$|\Lambda^{(i)}{}^0_{02}\rangle$. Finally, we should make
gauge transformation with the parameter
$|\Lambda^{(i+1)}{}^1_{00}\rangle$, removing
$|\Lambda^{(i)}{}^1_{02}\rangle$.
Thus, we have shown that the dependence on $\mathcal{P}_2^+$ can be
eliminated from $|\Lambda^{(i)}{}^j_0\rangle$.

This procedure works perfectly well until the terms linear in $p_1^+$
appear in the gauge functions $|\Lambda^{(i+1)}\rangle$.
When these terms are present, some of the gauge parameters
remain unused after eliminating the $\mathcal{P}_2^+$ dependence.
Due to the presence of $r$-dependent terms in
(\ref{dL02}), (\ref{dL12}), it is obvious that
if one should make a gauge transformation
with such a parameter the terms depending on the ghost
$\mathcal{P}_2^+$ may appear again.
Therefore, one should make gauge transformations with parameters
being linear in $p_1^+$ or independent of it,
before removing the $\mathcal{P}_2^+$ dependence.
The first gauge function where such a term appear is
$|\Lambda^{(1)}\rangle$.
Let us consider gauge transformation with this gauge
function more carefully.

Suppose that the dependence on the ghost $\mathcal{P}_2^+$ in
$|\Lambda^{(1)}\rangle$ has been removed by a gauge transformation.
Let us decompose the gauge functions
$|\Lambda^{(1)}\rangle$ and $|\Lambda^{(0)}\rangle$
as follows:
\begin{eqnarray}
|\Lambda^{(1)}{}^0_{0}\rangle
&=&
-ip_1^+\mathcal{P}_1^+|\omega\rangle+(p_1^+)^2(\ldots),
\\
|\Lambda^{(1)}{}^1_{0}\rangle&=&(p_1^+)^2(\ldots),
\\
|\Lambda^{(0)}{}^0_{0}\rangle
&=&
\mathcal{P}_1^+|\varepsilon\rangle
-i
p_1^+|\varepsilon_1\rangle
-i
\eta_1^+p_1^+\mathcal{P}_1^+|\varepsilon_2\rangle
\nonumber
\\
&&\qquad{}
-iq_1^+p_1^+\mathcal{P}_1^+|\varepsilon_3\rangle
-i
\eta_2^+p_1^+\mathcal{P}_1^+|\varepsilon_4\rangle
+(p_1^+)^2(\ldots)
+\mathcal{P}_2^+(\ldots)
,
\label{L00}
\\
|\Lambda^{(0)}{}^1_{0}\rangle
&=&
-ip_1^+\mathcal{P}_1^+|\varepsilon_5\rangle
+(p_1^+)^2(\ldots)
+\mathcal{P}_2^+(\ldots)
.
\label{L10}
\end{eqnarray}
As has been shown, we have to make a gauge transformation with
a parameter linear in $p_1^+$ (parameters which do not
depend on $p_1^+$ are absent from $|\Lambda^{(1)}\rangle$).
Therefore, we use $|\omega\rangle$ to make such a gauge transformation.
Since
\begin{eqnarray}
\delta|\varepsilon\rangle&=&-T_1^+|\omega\rangle
,
\end{eqnarray}
we use $|\omega\rangle$ to eliminate the dependence on
$b_2^+$ and $f^+$ from $|\varepsilon\rangle$,
\begin{eqnarray}
b_2|\varepsilon\rangle=f|\varepsilon\rangle=0\,,
\label{re}
\end{eqnarray}
and then we remove the $\mathcal{P}_2^+$ dependence from
$|\Lambda^{(0)}\rangle$ as described above.

Let us turn to the gauge-fixing of the fields.
We decompose the field in ghosts $\mathcal{P}_1^+$,
$\mathcal{P}_2^+$ by analogy with the gauge functions,
\begin{eqnarray}
|\chi^j_0\rangle
&=&
|\chi^j_{00}\rangle
+\mathcal{P}_1^+|\chi^j_{01}\rangle
+\mathcal{P}_2^+|\chi^j_{02}\rangle
+\mathcal{P}_1^+\mathcal{P}_2^+|\chi^j_{03}\rangle
,
\qquad
j=0,1,
\end{eqnarray}
with each $|\chi^j_{0m}\rangle$ being an expansion in $p_1^+$,
\begin{eqnarray}
|\chi^j_{0m}\rangle
&=&
\sum_{k=0} (p_1^+)^k |\chi^{jk}_{0m}\rangle
.
\end{eqnarray}
By analogy with the gauge-fixing of $|\Lambda^{(0)}\rangle$, we
have to use $|\varepsilon_i\rangle$ and then remove
the $\mathcal{P}_2^+$ dependence from the fields $|\chi^j_0\rangle$.

Substituting (\ref{L00}), (\ref{L10}) into (\ref{GT1}),
(\ref{GT2}), we find a gauge transformation for the fields
independent of the ghost $p_1^+$,
\begin{eqnarray}
\delta|\chi^{00}_{00}\rangle
&=&
L_1^+|\varepsilon\rangle+T_1^+|\varepsilon_1\rangle
,
\\
\delta|\chi^{00}_{01}\rangle
&=&
\eta_1^+\bigl(
-T_1^+|\varepsilon_2\rangle-L_1|\varepsilon\rangle-|\varepsilon_5\rangle
\bigr)
+\eta_2^+\bigl(
-T_1^+|\varepsilon_4\rangle-L_2|\varepsilon\rangle-|\varepsilon_3\rangle
\bigr)
\nonumber
\\
&&{}
+q_1^+\bigl(
-T_1^+|\varepsilon_3\rangle-T_1|\varepsilon\rangle
\bigr)
,
\\
\delta|\chi^{10}_{01}\rangle
&=&
-T_0|\varepsilon\rangle-2|\varepsilon_1\rangle-T_1^+|\varepsilon_5\rangle
.
\end{eqnarray}
Thus, we can first remove the $b_2^+$ and $f^+$ dependence of
$|\chi^{00}_{00}\rangle$, using $|\varepsilon_1\rangle$, and then
remove the $b_1^+$ dependence of $|\chi^{00}_{00}\rangle$, using
the restricted (\ref{re}) gauge parameter $|\varepsilon\rangle$.
We then remove the $b_2^+$ and $f^+$ dependence of
$|\chi^{10}_{01}\rangle$ and $|\chi^{00}_{01}\rangle$, using
the gauge parameters $|\varepsilon_5\rangle$,
$|\varepsilon_2\rangle$,
$|\varepsilon_3\rangle$, $|\varepsilon_4\rangle$.
After this, we remove the $\mathcal{P}_2^+$ dependence of the fields.
Now, all the gauge parameters have been used, and we have the following
conditions for the fields:
\begin{eqnarray}
&&
b_1|\chi^{00}_{00}\rangle=
b_2|\chi^{00}_{00}\rangle=
f|\chi^{00}_{00}\rangle=0
,
\label{gx0000}
\\
&&
b_2|\chi^{j0}_{01}\rangle=f|\chi^{j0}_{01}\rangle=0,
\qquad
j=0,1
\;,
\label{gxj001}
\\
&&
\eta_2|\chi^{0}_{0}\rangle=\eta_2|\chi^{1}_{0}\rangle=0.
\label{n2x}
\end{eqnarray}
Since $gh(|\chi^{00}_{00}\rangle)=0$, it does not depend on
ghost ``coordinates'' and due to (\ref{gx0000}) we conclude,
after gauge-fixing, that
\begin{eqnarray}
|\chi^{00}_{00}\rangle&=&|\Phi\rangle,
\label{PhSt}
\end{eqnarray}
where $|\Phi\rangle$ is the physical field (\ref{PhysState}).

Let us turn to removing the auxiliary fields with the help of
a part of the equations of motion.

\subsection{Removing auxiliary fields by equations
of motion}

Let us decompose equation (\ref{EofM1}) in $\mathcal{P}_1^+$,
$\mathcal{P}_2^+$ and consider the coefficients of $\mathcal{P}_2^+$-dependent
parts:
\begin{eqnarray}
-2\mathcal{P}_2^+\mathcal{P}_1^+\;:
&\quad&
q_1^2|\chi^0_{01}\rangle=0,
\label{0P12}
\\
-2\mathcal{P}_2^+\;:
&\quad&
q_1^2|\chi^0_{00}\rangle
-r\eta_1^+(L_2-2l_2')|\chi^0_{01}\rangle=0
.
\label{0P2}
\end{eqnarray}
Using (\ref{0P12}), we conclude that $|\chi^0_{01}\rangle$
contains no terms of an order larger than the first order in
$p_1^+$,
\begin{eqnarray}
|\chi^0_{01}\rangle
&=&
|\chi^{00}_{01}\rangle+p_1^+|\chi^{01}_{01}\rangle
.
\label{x001}
\end{eqnarray}
Substituting (\ref{x001}) to (\ref{0P2}), we find
\begin{eqnarray}
\label{x000} |\chi^0_{00}\rangle
&=&\sum_{k=0}^3(p_1^+)^k|\chi^{0k}_{00}\rangle,
\end{eqnarray}
where
\begin{eqnarray}
\label{x0200}
|\chi^{02}_{00}\rangle
=
-\frac{r}{2}\;\eta_1^+(L_2-2l_2')|\chi^{00}_{01}\rangle
,
\qquad
|\chi^{03}_{00}\rangle
=
-\frac{r}{6}\;\eta_1^+(L_2-2l_2')|\chi^{01}_{01}\rangle
.
\end{eqnarray}

In a similar way, we decompose equation (\ref{EofM2}) and obtain
\begin{eqnarray}
-2\mathcal{P}_2^+\mathcal{P}_1^+\;:
&\quad&
q_1^2|\chi^1_{01}\rangle=0,
\label{1P12}
\\
-2\mathcal{P}_2^+\;:
&\quad&
q_1^2|\chi^1_{00}\rangle
-r\eta_1^+(L_2-2l_2')|\chi^1_{01}\rangle
-\frac{r}{2}(T_1-2t_1')|\chi^0_{01}\rangle
=0
.
\label{1P2}
\end{eqnarray}
These equations have the following solution:\footnote{The term
$|\chi^{10}_{00}\rangle$ corresponding to $(p_1^+)^0$
is absent in the sum of (\ref{x101}) due
to the ghost number restriction (\ref{0chi}).}
\begin{eqnarray}
\label{x101} |\chi^1_{01}\rangle =
|\chi^{10}_{01}\rangle+p_1^+|\chi^{11}_{01}\rangle , \qquad
|\chi^1_{00}\rangle =\sum_{k=1}^3(p_1^+)^k|\chi^{1k}_{00}\rangle,
\end{eqnarray}
where
\begin{eqnarray}
\label{x1200}
|\chi^{12}_{00}\rangle
&=&
-\frac{r}{2}\;\eta_1^+(L_2-2l_2')|\chi^{10}_{01}\rangle
-\frac{r}{4}\;(T_1-2t_1')|\chi^{00}_{01}\rangle
,
\\
\label{x1300}
|\chi^{13}_{00}\rangle
&=&
-\frac{r}{6}\;\eta_1^+(L_2-2l_2')|\chi^{11}_{01}\rangle
-\frac{r}{12}\;(T_1-2t_1')|\chi^{01}_{01}\rangle
.
\end{eqnarray}

Let us now consider a part of equations (\ref{EofM1}),
(\ref{EofM2}) containing the physical field
$|\chi^{00}_{00}\rangle=|\Phi\rangle$ (\ref{PhSt}),
\begin{eqnarray}
&&
(t_0+\tilde{\gamma}m_0)|\Phi\rangle
+iT_1^+|\chi^{11}_{00}\rangle
+L_1^+|\chi^{10}_{01}\rangle
-r(L_2^+-2l_2^{\prime+})|\chi^{001}_{010}\rangle
=0,
\label{1}
\\
q_1^+\;:
&&
t_1|\Phi\rangle+iT_1^+|\chi^{011}_{000}\rangle
+L_1^+|\chi^{001}_{010}\rangle-|\chi^{10}_{01}\rangle
=0,
\label{q1+}
\\
\eta_1^+\;:
&&
l_1|\Phi\rangle-iT_1^+|\chi^{010}_{001}\rangle
+L_1^+|\chi^{000}_{011}\rangle-i|\chi^{11}_{00}\rangle
\nonumber
\\
&&\qquad{}
-2r(L_2^+-2l_2^{\prime+})|\chi^{000}_{012}\rangle
+\frac{3r}{4}\;(T_1^+-2t_1^{\prime+})|\chi^{001}_{010}\rangle
-T_0|\chi^{10}_{01}\rangle
=0,
\label{n1+}
\\
\eta_2^+\;:
&&
l_2|\Phi\rangle-iT_1^+|\chi^{010}_{002}\rangle+i|\chi^{011}_{000}\rangle
+L_1^+|\chi^{000}_{012}\rangle-|\chi^{000}_{011}\rangle
=0
,
\label{n2+}
\end{eqnarray}
where we have decomposed the fields $|\chi^{00}_{01}\rangle$ and
$|\chi^{01}_{00}\rangle$ as follows:
\begin{eqnarray}
\label{x0001d} |\chi^{00}_{01}\rangle=
q_1^+|\chi^{001}_{010}\rangle +\eta_1^+|\chi^{000}_{011}\rangle
+\eta_2^+|\chi^{000}_{012}\rangle , \qquad |\chi^{01}_{00}\rangle=
q_1^+|\chi^{011}_{000}\rangle +\eta_1^+|\chi^{010}_{001}\rangle
+\eta_2^+|\chi^{010}_{002}\rangle \,.
\end{eqnarray}

Acting on (\ref{q1+}) by the operators $f$ and $b_2$ and taking into
account the gauge-fixing condition (\ref{gx0000}), (\ref{gxj001}),
we obtain, respectively,
\begin{eqnarray}
fT_1^+|\chi^{011}_{000}\rangle=0, \qquad
b_2T_1^+|\chi^{011}_{000}\rangle=0.
\end{eqnarray}
These equations yield
\begin{eqnarray}
|\chi^{011}_{000}\rangle=0
.
\label{011000}
\end{eqnarray}
Acting on (\ref{n2+}) by $f$ and $b_2$, we obtain
\begin{eqnarray}
fT_1^+|\chi^{010}_{002}\rangle=
b_2T_1^+|\chi^{010}_{002}\rangle=
0
&\Longrightarrow&
|\chi^{010}_{002}\rangle=0
.
\label{010002}
\end{eqnarray}

We now act on equation (\ref{1}) by the operator $f$,
\begin{eqnarray}
fT_1^+|\chi^{11}_{00}\rangle=0,
\end{eqnarray}
and then presenting the vector $|\chi^{11}_{00}\rangle$ as
a power series in $f^+$
\begin{eqnarray}
|\chi^{11}_{00}\rangle
&=&
|a^{11}_{00}\rangle+f^+|b^{11}_{00}\rangle,
\end{eqnarray}
we have
\begin{eqnarray}
|a^{11}_{00}\rangle&=&t_1^+|b^{11}_{00}\rangle
.
\end{eqnarray}
Substituting this result into equation (\ref{1}) and acting
twice by the operator $b_2$ on the resulting equations, we
arrive at
\begin{eqnarray}
b_2|b^{11}_{00}\rangle&=&0.
\end{eqnarray}
Finally, equation (\ref{1}) acquires the form
\begin{eqnarray}
&&
(t_0+\tilde{\gamma}m_0)|\Phi\rangle
+2iL_2^+|b^{11}_{00}\rangle
+L_1^+|\chi^{10}_{01}\rangle
-r(L_2^+-2l_2^{\prime+})|\chi^{001}_{010}\rangle
=0.
\label{1'}
\end{eqnarray}

Let us present the state vectors $|b^{11}_{00}\rangle$,
$|\chi^{10}_{01}\rangle$, $|\chi^{001}_{010}\rangle$ as
a power series in $b_1^+$,
\begin{align}
&
|b^{11}_{00}\rangle=\sum_{k=0}^{n-2}(b_1^+)^k|b^{11}_{00}\rangle^k
, &&
|\chi^{10}_{01}\rangle=\sum_{k=0}^{n-1}(b_1^+)^k|\chi^{10}_{01}\rangle^k
, &&
|\chi^{001}_{010}\rangle=\sum_{k=0}^{n-2}(b_1^+)^k|\chi^{001}_{010}\rangle^k
,
\end{align}
and then, presenting equations (\ref{1'}) and (\ref{q1+}) as power
series in $b_1^+$ and $b_2^+$, we have
\begin{eqnarray}
\label{1'n}
(b_1^+)^n:
&&
m_1|\chi^{10}_{01}\rangle^{n-1}=0,
\\
\label{1'n-1}
(b_1^+)^{n{-}1}:
&&
m_1|\chi^{10}_{01}\rangle^{n-2}=-l_1^+|\chi^{10}_{01}\rangle^{n-1},
\\
\label{1'k}
1\le k \le n{-}2
\quad
(b_1^+)^k:
&&
m_1|\chi^{10}_{01}\rangle^{k-1}
=
-l_1^+|\chi^{10}_{01}\rangle^k
-2il_2^+|b^{11}_{00}\rangle^k
+rl_2^+|\chi^{001}_{010}\rangle^k
,
\\
\label{1'0}
(b_1^+)^0:
&&
(t_0+\tilde{\gamma}m_0)|\Phi\rangle
=
-2il_2^+|b^{11}_{00}\rangle^0
-l_1^+|\chi^{10}_{01}\rangle^0
+rl_2^+|\chi^{001}_{010}\rangle^0
,
\end{eqnarray}
\begin{eqnarray}
\label{1'b2} 0\le k \le n{-}2 \qquad b_2^+(b_1^+)^k: &\qquad&
|b^{11}_{00}\rangle^k=\frac{ir}{2}\,|\chi^{001}_{010}\rangle^k ,
\end{eqnarray}
\begin{eqnarray}
\label{q1+n-1}
(b_1^+)^{n-1}:
&&
m_1|\chi^{001}_{010}\rangle^{n-2}
=
|\chi^{10}_{01}\rangle^{n-1}
,
\\
\label{q1+k}
1\le k\le n{-2}
\quad
(b_1^+)^k:
&&
m_1|\chi^{001}_{010}\rangle^{k-1}
=
|\chi^{10}_{01}\rangle^k
-l_1^+|\chi^{001}_{010}\rangle^k
,
\\
\label{q1+0}
(b_1^+)^0:
&&
t_1|\Phi\rangle
=
|\chi^{10}_{01}\rangle^0
-l_1^+|\chi^{001}_{010}\rangle^0.
\end{eqnarray}
Using (\ref{1'n}) and (\ref{1'n-1}), we have
$|\chi^{10}_{01}\rangle^{n-1}=|\chi^{10}_{01}\rangle^{n-2}=0$.
Substituting this result into (\ref{q1+n-1}) and then into
(\ref{q1+k}) for $k=n-2$, we obtain
$|\chi^{001}_{010}\rangle^{n-2}=|\chi^{001}_{010}\rangle^{n-3}=0$.
Turning to equation (\ref{1'b2}) for $k=n-2$ and $k=n-3$, we
conclude that
$|b^{11}_{00}\rangle^{n-2}=|b^{11}_{00}\rangle^{n-3}=0.$ We now
repeat the procedure starting from equation (\ref{1'k}) for
$k=n-2$ and $k=n-3$. Finally, we obtain
\begin{eqnarray}
\label{TheEq'}
&&
(t_0+\tilde{\gamma}m_0)|\Phi\rangle
=0,
\qquad
t_1|\Phi\rangle
=0,
\\
&&
|\chi^{10}_{01}\rangle
=|\chi^{001}_{010}\rangle=0
,
\qquad
|b^{11}_{00}\rangle=0
\quad
\Longrightarrow
|\chi^{11}_{00}\rangle=0
.
\label{b}
\end{eqnarray}
Using (\ref{TheEq'}), we can see that the physical state
(\ref{PhysState}) satisfies (\ref{Eq-0}), (\ref{Eq-1}), or equivalently
(\ref{TheEq}), provided that condition (\ref{final m0}) is taken into account.

Let us now turn to equations (\ref{n1+}), (\ref{n2+}).
Taking into account (\ref{011000}), (\ref{010002}), (\ref{b}),
we can see that these two equations read
\begin{eqnarray}
&&
l_1|\Phi\rangle
-iT_1^+|\chi^{010}_{001}\rangle
+L_1^+|\chi^{000}_{011}\rangle
-2r(L_2^+-2l_2^{\prime+})|\chi^{000}_{012}\rangle
=0,
\label{n1}
\\
&&
l_2|\Phi\rangle
+L_1^+|\chi^{000}_{012}\rangle
-|\chi^{000}_{011}\rangle
=0
.
\label{n2}
\end{eqnarray}
They are analogous to (\ref{1}) and (\ref{q1+}), where
$|\chi^{011}_{000}\rangle=0$ in (\ref{011000}) is taken into
account. Therefore, we can repeat the procedure that has been
carried out with equations (\ref{1}), (\ref{q1+}), and thus we
arrive at the conclusion that
\begin{eqnarray}
\label{l1l2}
&&
l_1|\Phi\rangle=l_2|\Phi\rangle=0,
\\
&&
|\chi^{010}_{001}\rangle=
|\chi^{000}_{012}\rangle=
|\chi^{000}_{011}\rangle=0.
\label{010001}
\end{eqnarray}
Equations (\ref{l1l2}) are consequences of (\ref{TheEq'}); they do
not impose any additional restrictions on the physical state
(\ref{PhysState}).
Collecting (\ref{011000}), (\ref{010002}), (\ref{b}),
(\ref{010001}) and (\ref{x0001d}), we have
\begin{eqnarray}
|\chi^{10}_{01}\rangle=
|\chi^{11}_{00}\rangle=
|\chi^{00}_{01}\rangle=
|\chi^{01}_{00}\rangle=0
\label{a1=0}
.
\end{eqnarray}
Observing (\ref{x001}), (\ref{x000}), (\ref{x0200}) and
(\ref{x101}), (\ref{x1200}), (\ref{x1300}), we can see that
it remains to show that
$|\chi^{01}_{01}\rangle=|\chi^{11}_{01}\rangle=0$.
To prove this fact, we decompose (\ref{EofM1}) and (\ref{EofM2}) in
the ghosts $p_1^+$, $\mathcal{P}_1^+$ and consider the equations which
are the coefficients at $(p_1^+)^0\mathcal{P}_1^+$. With allowance for
(\ref{a1=0}), these equations read
\begin{eqnarray}
iT_1^+|\chi^{01}_{01}\rangle -\eta_2^+p_1|\chi^{01}_{01}\rangle=0,
\qquad iT_1^+|\chi^{11}_{01}\rangle
-\eta_2^+p_1|\chi^{11}_{01}\rangle=0\,, \label{x-x}
\end{eqnarray}
respectively.
Decomposing $|\chi^{01}_{01}\rangle$, $|\chi^{11}_{01}\rangle$
in ghost ``coordinates'',
\begin{eqnarray}
|\chi^{01}_{01}\rangle
=
q_1^+|\chi^{011}_{010}\rangle
+\eta_1^+|\chi^{010}_{011}\rangle
+\eta_2^+|\chi^{010}_{012}\rangle
,
\qquad
|\chi^{11}_{01}\rangle
=
q_1^+|\chi^{111}_{010}\rangle
+\eta_1^+|\chi^{110}_{011}\rangle
+\eta_2^+|\chi^{110}_{012}\rangle\,,
\label{0101d}
\end{eqnarray}
and substituting  the result into (\ref{x-x}), we obtain equations
for the coefficients of (\ref{0101d}) in the form
\begin{eqnarray}
T_1^+|\chi^{j1i}_{01n}\rangle=0 \quad \Longrightarrow \quad
|\chi^{j1i}_{01n}\rangle=0 .
\end{eqnarray}
Therefore, all the coefficient in decompositions (\ref{0101d}) of
$|\chi^{01}_{01}\rangle$, $|\chi^{11}_{01}\rangle$ are equal to zero,
and we conclude that
\begin{eqnarray}
\label{a2=0}
|\chi^{01}_{01}\rangle=|\chi^{11}_{01}\rangle=0
.
\end{eqnarray}
Thus, we have shown that after the gauge fixing
(\ref{gx0000})--(\ref{n2x}) all the auxiliary fields become equal to zero,
(\ref{a1=0}), (\ref{a2=0}), and the physical state
$|\Phi\rangle=|\chi^{00}_{00}\rangle$ (\ref{PhSt}) obeys
equations (\ref{Eq-0}), (\ref{Eq-1}).

Let us now consider some examples of the Lagrangian construction procedure.

\section{Examples}\label{examples}

Here, we shall illustrate the general procedure of gauge-invariant
Lagrangian construction by using the examples of fermionic fields
of spin $1/2$ and $3/2$.

\subsection{Spin-1/2 field}
For a fermionic field of spin $s=\frac{1}{2}$, we have $h =
2-\frac{d}{2}$. Then the only nonvanishing vector
$|\chi^0_0\rangle_0$ subject to condition (\ref{schi}) and
having the proper ghost number (\ref{Qchi}) has the form
\begin{eqnarray}
|\chi^0_0\rangle_0 = \psi(x)|0\rangle,  \qquad
{}_0\langle\tilde{\chi}^0_0| = {}_0\langle
0|{\psi}^+(x)\tilde{\gamma}{}^0.
\end{eqnarray}
Then, due to (\ref{K}), (\ref{tK}) for $K_0
=|0\rangle\langle0|+\ldots$,
(\ref{final m0}), the action implied by (\ref{L1}) has the form
\begin{eqnarray}
\label{s12} \mathcal{S}_0 &=& {}_0\langle\tilde{\chi}_0^0|K_0
  T_0|\chi_0^0\rangle_0 = -\int d^dx \sqrt{|g|}\bar{\psi}\Bigl\{ i \gamma^{\mu}\nabla_{\mu} - m -
  r^{\frac{1}{2}}\Bigl(\frac{d}{2}-2  \Bigr)\Bigr\}\psi.
\end{eqnarray}
Here, we have applied the definition (\ref{gammas}) for the conventional
gamma-matrices and have introduced the Dirac-conjugate spinor
$\bar{\psi}$, $\bar{\psi}=\psi^+ \gamma^0$. Thus, we can see that
the action  (\ref{s12}) reproduces equation (\ref{Eq-0})
for $n=0$, which corresponds to spin-1/2 field.

\subsection{Spin-3/2 field}

In the case of a spin-3/2 field, we have $n=1$, $h=1-d/2$,
$m_0=m+r^{\frac{1}{2}}(d-2)/2$. Since $i_{max}=0$, the
corresponding Lagrangian formulation is an irreducible gauge
theory. Due to ${gh}(|\Lambda^1_0\rangle_1) = -2$), the
nonvanishing fields $|\chi^0_0\rangle_1$, $|\chi^1_0\rangle_1$ and
the gauge parameter $|\Lambda^0_0\rangle_1$, for
$|\Lambda^1_0\rangle_1 \equiv 0$, possess the following Grassmann
grading and ghost number distributions:
\begin{eqnarray}
& \left(\varepsilon, {gh}\right)(|\chi_0^0\rangle_1)=(1, 0),\quad
\left(\varepsilon, {gh}\right)(|\chi_0^1\rangle_1)=(1, -1),\quad
\left(\varepsilon, {gh}\right)(|\Lambda_0^0\rangle_1)=(0,-1).
&
\end{eqnarray}
These conditions determine the dependence of the fields and gauge
parameter on the oscillator variables in a unique form:
\begin{align}\label{chi32}
& |\chi_0^0\rangle_1 = \left[ -ia^{+\mu}\psi_{\mu}(x) + f^+
\tilde{\gamma}\psi(x) + b_1^+\varphi(x) \right]|0\rangle, &&
|\chi_0^1\rangle_1 = \left[ \mathcal{P}^{+}_1\tilde{\gamma}\chi(x)
+i p_1^+\chi_1(x)
\right]|0\rangle,\\
& \label{brachi32} {}_1\langle\tilde{\chi}_0^0| = \langle0|\left[
i\psi_{\mu}^+(x)a^{\mu} +  \psi^+(x)\tilde{\gamma}f+
\varphi^+(x)b_1 \right]\tilde{\gamma}^0, &&
{}_1\langle\tilde{\chi}_0^1| = \langle 0|
\left[\chi^+(x)\tilde{\gamma}\mathcal{P}_1-
    \chi^+_1ip_1
\right]\tilde{\gamma}^0, \\
&
|\Lambda_0^0\rangle_1
=
\left[
\mathcal{P}^{+}_1\xi_1(x)-ip_1^+\tilde{\gamma}\xi_2(x)
\right]|0\rangle
.
\label{l001}
\end{align}
Substituting (\ref{chi32}), (\ref{brachi32}) into (\ref{L1}), we
find the action   (up to an overall factor) for a spin-$3/2$ field
interacting with the AdS background:
\begin{eqnarray}
{\cal{}S}_1 &=& \int d^dx \sqrt{|g|}\left[\bar{\psi}^\mu\Bigl\{
\bigl[i\gamma^\nu\nabla_\nu-m_0\bigr]\psi_\mu -\nabla_\mu\chi
+i\gamma_\mu\chi_1 \Bigr\} \right.\nonumber
\\
&&{}
+
\Bigl[
(d-2)\bar{\psi}
-
\frac{m_0}{m_1}\bar{\varphi}
\Bigr]
\Bigl\{
\bigl[i\gamma^\mu\nabla_\mu+m_0\bigr]\psi
-\frac{r(d-1)}{2m_1}\varphi
-\chi_1
\Bigr\}
\nonumber
\\
&&{}
-
\Bigl[
\frac{M^2}{m_1^2}\bar{\varphi}
+\frac{m_0}{m_1}\bar{\psi}
\Bigr]
\Bigl\{
\bigl[i\gamma^\sigma\nabla_\sigma-m_0\bigr]\varphi
-2m_1\psi
-m_1\chi
\Bigr\}
\nonumber
\\
&&{}
+
\bar{\chi}\Bigl\{
\bigl[
i\gamma^\mu\nabla_\mu
+m_0
\bigr]
\chi
+\chi_1
+\nabla^\mu\psi_\mu
+m_0\psi
+
\frac{M^2}{m_1}\;
\varphi
\bigr\}
\nonumber
\\
&&{} \left.-\bar{\chi}_1\Bigl\{ i\gamma^\mu\psi_\mu +(d-2)\psi
-\chi -\frac{m_0}{m_1}\varphi \Bigr\}\right] \label{L3/2} ,
\end{eqnarray}
where $M^2=m_0^2-\frac{1}{4}\;r(d-1)(d-2)$. To obtain the action
(\ref{L3/2}), we have used the expressions for the operators $K_1$
(\ref{K}), (\ref{tK}). Substituting (\ref{chi32})--(\ref{l001})
into (\ref{GT1}), (\ref{GT2}), we find the gauge transformations
\begin{eqnarray}
&&
\delta\psi_\mu
=\nabla_\mu\xi_1+i\gamma_\mu\xi_2,
\hspace*{4em}
\delta\psi=\xi_2,
\hspace*{4em}
\delta\varphi=m_1\xi_1
\\
&&
\delta\chi=
\Bigl[
i\gamma^\mu\nabla_\mu
-m_0
\Bigr]\xi_1
-2\xi_2,
\hspace*{4em}
\delta\chi_1
=
\Bigl[
i\gamma^\mu\nabla_\mu
+m_0
\Bigr]
\xi_2
-\frac{r}{2}(d-1)
\xi_1
.
\label{gtr}
\end{eqnarray}

Let us present the action in terms of one physical field
$\psi_\mu$.
To this end, we get rid of the fields $\varphi$,
$\psi$, by using their gauge transformations and the
gauge parameters $\xi_1$, $\xi_2$, respectively. Having expressed
the field $\chi$, using the equation of motion $\chi=i\gamma^\mu\psi_\mu$,
we can see that the terms with the Lagrangian multiplier $\chi_1$ turn to zero.
As a result, we obtain \begin{eqnarray}
\mathcal{L}_{RS} &=&
\bar{\psi}^\mu(i\gamma^\sigma\nabla_\sigma-m_0)\psi_\mu
-i\bar{\psi}^\mu(\gamma_\nu\nabla_\mu+\gamma_\mu\nabla_\nu)\psi^\nu
+\bar{\psi}^\nu\gamma_\nu(i\gamma^\sigma\nabla_\sigma+m_0)\gamma^\mu\psi_\mu
.
\end{eqnarray}
This is a generalization of the Rarita--Schwinger Lagrangian to
a $d$-dimensional AdS space.

\section{Conclusion}\label{Conclusion}

We have constructed a gauge-invariant Lagrangian formulation of
half-integer totally symmetric higher spin fields in the AdS space
of any dimension in the ``metric-like" formulation. The results of
this study are most general and apply to both massive and massless
fermionic higher spin fields in the AdS, Minkowski, and
dS\footnote{In the case of HS fields on dS space the corresponding
Lagrangians lose the property of reality  due to
$r^{\frac{1}{2}}$.} spaces.

Starting from embedding the fermionic higher spin fields
into vectors of an auxiliary Fock space, we treat the fields
as components of these Fock-space vectors, and, as a result, we
reformulate the theory in terms of such vectors. We realize the
conditions that define an irreducible representation of the AdS group
with a given mass and spin in terms of  differential operators
acting in this Fock space. The mentioned conditions are
interpreted as constraints imposed on the Fock space vectors and generate
a closed higher spin nonlinear symmetry superalgebra being the
basic object of this study.

It is shown that the derivation  of a correct Lagrangian
formulation requires a  transition to another constraint basis for
the original symmetry algebra, which is algebraically equivalent
to the initial basis and does not contain the constraints that
define an irreducible representation of the AdS group. This set of
constraints provides an additive extension of the initial algebra
to another operator algebra. Then, as shown in \cite{0608005}, it
is necessary to deform the initial algebra. As a result, one
obtains a nonlinear superalgebra of enlarged constraints, whose
construction by means of the additional operators realizes a
special conversion of the initial system of first- and
second-class operator constraints into a system of first-class
ones with a preservation of the initial algebraic structure for
the deformed constraints (see \cite{conversion} for the
elaboration of conversion methods).
 Due to the nonlinearity of
the underlying  algebra of the enlarged constraints, the
corresponding  BRST operator is defined ambiguously, and we
construct it in an exact form with a non-vanishing term of third
order in powers of ghosts for the case of a supersymmetric
ordering of the constraints in  commutator relations. It is shown
that the resulting BRST operator yields a consistent Lagrangian
dynamics for fermionic fields of any spin. The corresponding
Lagrangian formulation is constructed in a concise form in terms
of a Fock space and proves to be a reducible gauge theory with a
finite number of reducibility stages, the number growing with the
spin value. It is interesting to observe that the methods
developed for the quantization of gauge theories turn out to be
extremely efficient for deriving classical gauge-invariant
Lagrangians for higher spin field theories, thus reflecting one
more side of  the BV--BFV duality concept
\cite{GrigorievDamgaard}, which permits one to construct, by means
of a Hamiltonian BFV--BRST charge, the objects used in Lagrangian
formalism.

We have proved that the Lagrangian equations of motion
(\ref{EofM1}), (\ref{EofM2}), after a partial gauge-fixing,
 reproduce the equations corresponding
to the relations determining the irreducible representation of the
AdS group. This proof completes the derivation of a
correct gauge-invariant Lagrangian formulation for  higher spin
fermionic fields in the AdS space. As examples demonstrating the
general scheme of Lagrangian construction, we have obtained
gauge-invariant Lagrangian descriptions of fields with spin-$1/2$ and
spin-$3/2$ in an explicit form. In principle, the
constructed description permits one to obtain explicit
Lagrangians for any other half-integer-spin fields.

The basic results of the present work are given by relations
(\ref{L1}), where the Lagrangian action for a field with an
arbitrary half-integer spin is constructed, and by
(\ref{GT1})--(\ref{GTi2}), where the gauge transformations for the
fields are presented as well as the sequence of reducible gauge
transformations for the gauge parameters.

Concluding, we shall discuss two points. First, the
gauge-invariant formulation for massive higher spin field theories
in the AdS space fields can have very interesting applications in
the calculation of a quantum effective action for higher spin
massive fields in the AdS space. The Lagrangians of all such
models have a gauge-invariant kinetic term and a mass term
violating the gauge invariance. Such a structure of a Lagrangian
leads to some problems of quantum calculations in a curved
space-time. To avoid these problems, it is natural to construct a
gauge-invariant formulation by using appropriate St$\ddot{\rm
u}$ckelberg fields. Then, one can impose gauge-fixing conditions
removing the gauge degeneration of the kinetic term in the
Lagrangian of the basic fields and apply the standard techniques
for the calculation of an effective action (see the examples of
such calculations in \cite{0703189}). We emphasize that the BRST
approach leads automatically to a gauge-invariant Lagrangian with
the entire set of appropriate St$\ddot{\rm u}$ckelberg fields. As
a result, one has a basis for constructing the effective action of
massive higher spin fields in the AdS space. Second, the BRST
approach can provide a systematic method of constructing
interaction vertices for higher spin fields in the AdS space
\cite{0609082}. Therefore, one can hope that the BRST construction
developed in this paper will be useful in deriving interaction
vertices for massive  higher spin fermionic fields in the AdS
space.


\section*{Acknowledgements}
A.A.R. is
grateful to M. Grigoriev and P. Moshin for useful discussions. The work of I.L.B
and V.A.K was partially supported by the INTAS grant, project
INTAS-05-7928, the RFBR grant, project No.\ 06-02-16346, and the
grant for LRSS, project No.\ 4489.2006.2. The work of I.L.B was
supported in part by the DFG grant, project No.\ 436 RUS
113/669/0-3, and the joint RFBR-DFG grant, project No.\
06-02-04012. The work of V.A.K was partially supported by the
joint DAAD--Mikhail Lomonosov Programme (Referat 325, Kennziffer
A/06/16774).


\end{document}